\documentclass[useAMS,usenatbib]{mn2e}

\usepackage{rotating}
\usepackage{graphicx}
\usepackage{txfonts}
\usepackage{lscape}
\usepackage{natbib}
\bibpunct{(}{)}{;}{a}{}{,} % to follow the A&A style

\title[Simulations of MOAO-fed IFS]{Coupling MOAO with Integral Field
  Spectroscopy: specifications for the VLT and the E-ELT} \author[M.
  Puech et al.]{M. Puech$^{1,2}$\thanks{E-mail: mpuech@eso.org}, H.
  Flores$^{2}$, M. Lehnert$^{2}$, B. Neichel$^{2,3}$, T. Fusco$^{3}$,
  P. Rosati$^{1}$, J.-G. Cuby$^{4}$,\newauthor and G. Rousset$^{5}$\\ $^{1}$ESO,
  Karl-Schwarzschild-Strasse 2, D-85748 Garching bei M\"unchen,
  Germany\\ $^{2}$GEPI, Observatoire de Paris, CNRS, University Paris
  Diderot; 5 Place Jules Janssen, 92190 Meudon, France\\ $^{3}$ONERA,
  BP 72, 92322 Chatillon Cedex, France\\ $^{4}$Laboratoire
  d'Astrophysique de Marseille, Observatoire Astronomique de
  Marseille-Provence, 2 Place Le Verrier, 13248 Marseille,
  France\\ $^{5}$LESIA, Observatoire de Paris, CNRS, UPMC, University
  Paris Diderot, 5 place Jules Janssen 92190 Meudon, France.\\}
\begin{document}

\date{Accepted ...}

\pagerange{\pageref{firstpage}--\pageref{lastpage}} \pubyear{2002}

\maketitle

\label{firstpage}

\begin{abstract}
Elucidating the processes that governed the assembly and evolution of
galaxies over cosmic time is one of the main objectives of all of the
proposed Extremely Large Telescopes (ELT). To make a leap forward in
our understanding of these processes, an ELT will want to take
advantage of Multi-Objects Adaptive Optics (MOAO) systems, which can
substantially improve the natural seeing over a wide field of view. We
have developed an end-to-end simulation to specify the science
requirements of a MOAO-fed integral field spectrograph on either an 8m
or 42m telescope. Our simulations re-scales observations of local
galaxies or results from numerical simulations of disk or interacting
galaxies. The code is flexible in that it allows us to explore a wide
range of instrumental parameters such as encircled energy (EE), pixel
size, spectral resolution, etc. For the current analysis, we limit
ourselves to a local disk galaxy which exhibits simple rotation and a
simulation of a merger. While the number of simulations is limited, we
have attempted to generalize our results by introducing the simple
concepts of ``PSF contrast'' which is the amount of light polluting
adjacent spectra which we find drives the smallest EE at a given
spatial scale. The choice of the spatial sampling is driven by the
``scale-coupling''. By scale-coupling we mean the relationship between
the IFU pixel scale and the size of the features that need to be
recovered by 3D spectroscopy in order to understand the nature of the
galaxy and its substructure. Because the dynamical nature of galaxies
are mostly reflected in their large-scale motions, a relatively coarse
spatial resolution is enough to distinguish between a rotating disk
and a major merger. Although we used a limited number of
morpho-kinematic cases, our simulations suggest that, on a 42m
telescope, the choice of an IFU pixel scale of 50-75 mas seems to be
sufficient. Such a coarse sampling has the benefit of lowering the
exposure time to reach a specific signal-to-noise as well as relaxing
the performance of the MOAO system. On the other hand, recovering the
full 2D-kinematics of z$\sim$4 galaxies requires high signal-to-noise
and at least an EE of 34\% in 150 mas (2 pixels of 75 mas). Finally,
we carried out a similar study for a hypothetical galaxy/merger at
z=1.6 with a MOAO-fed spectrograph for an 8m, and find that at least
an EE of 30\% at 0.25 arcsec spatial sampling is required to
understand the nature of disks and mergers.

\end{abstract}


\section{Introduction}
Developing a coherent model for the mass assembly of galaxies over
cosmic time is a complex and difficult task. It is difficult because
developing such a coherent model involves highly non-linear physics
(e.g., the cooling and collapse of baryons, or feedback from stars and
super-massive blackholes which regulate both the growth of the stellar mass
and black holes themselves) and all over a very wide range of physical
scales (from large scale structure to star clusters to black holes). Our
knowledge of galaxy formation  {\it in situ} relies mostly on observations
of integrated quantities,which is insufficient to understand the details
of the process of formation, e.g., the interplay between the baryonic
and dark matter, or the complex physics of baryons such as merging,
star formation, feedback, various instabilities and heating/cooling
mechanisms, etc. At the heart of our developing understanding of
galaxy evolution is the relative importance of secular evolution through
(quasi-)adiabatic accretion of mass, instabilities, and resonances (e.g.,
\citealt{Semelin05}), versus more violent evolution through merging
(e.g., \citealt{Hammer05}), as a function of look-back time.

Because of the complexity of the processes involved, understanding the
mass assembly of galaxies over cosmic time requires constraining a wide
range of phenomenology for large samples of objects. In this respect
integral field spectroscopy is particularly well suited in that it allows
us to map the spatially resolved physical properties of galaxies (see,
e.g., \citealt{Puech06b}). Unfortunately, splitting the light into many
channels in a integral field spectrograph requires long integration times,
even on large telescopes, to detect low surface brightness emission.
Thus, conducting large statistical studies necessary for understanding
the processes driving galaxy evolution will demand a high multiplex
capability to efficiently investigate sufficient numbers of objects.

Using current facilities on 8m class telescopes, it is already
possible to map the physical properties of distant galaxies. For
example, FLAMES/GIRAFFE on the VLT offers the unique ability to
observe 15 distant galaxies simultaneously. The first (complete) 3D
sample of galaxies at moderate redshifts has revealed a large fraction
($\sim$40\%) of all z$\sim$0.6 galaxies with $M_B <$-19.5 have
perturbed or complex kinematics (\citealt{Flores06, Yang07}). This
population of dynamically disturbed galaxies is most probably a result
of a high prevalence of mergers and merger remnants (see also
\citealt{Puech06a, Puech08}). At higher redshift, integral field
spectroscopy of several objects has been obtained using single object
integral field spectrographs available on 8-10 meter class telescopes
(e.g., \citealt{Forster06}). However, many of these data sets were
obtained with limited spatial resolution sampling galaxies on scales
of a few kpc, and the true dynamical nature of many sources sampled at
these spatial and spectral resolutions remains uncertain (e.g.,
\citealt{Law07}).

Improving and extending this approach to the general understanding of
the nature of faint galaxies and to galaxies in the early Universe
will require a new generation of multi-object integral field
spectrographs working in the near-IR, with increased sensitivity and
better spatial resolution on even larger telescopes. Because of the
complex interplay between spatial and spectral features and our
limited knowledge of high redshift galaxies, it is helpful, perhaps
necessary, to rely on numerical simulations for constraining the
design parameters of this new kind of instrument. For this purpose, we
have developed software that simulate end-to-end the emission line
characteristics of local galaxies and numerical simulations of
galaxies to show how they would appear in the distant Universe. By
changing various parameters like the resolution, pixel scale, and
point spread function it is possible to constrain the instrumental
characteristics and performance against a set of galaxy
characteristics (e.g., velocity field). This paper is the first in a
series to investigate the scientific design requirements of possible
future instruments, telescope aperture and design, and site
characteristics appropriate for constrain how galaxies grew and
evolved with cosmic time. In this paper in particular, we aim to
present the details of our assumptions that go into the software, as
well as a first attempt to constrain some of the high level
specification for a hypothetical but realistic multi-object integral
field spectrograph assisted by MOAO system. Because 2D kinematics
(i.e., velocity field and velocity dispersion maps) are one of the
most obvious and easiest to currently simulate, they provide the most
straight forward way of constraining the high level science design
requirements for this type of instrument. We emphasize that the goal
of this paper is not to study the detailed scientific capability of
such instruments. Instead, the present paper aims at examining a few
scientifically motivated cases to derive relevant specifications for
such types of instruments with a capable MOAO system. These
specifications will then be adopted as a baseline for studying their
scientific capabilities subsequent papers. This paper is organized as
follows: in Section 2, we detail the goals of this study and our
methodology; in Sect. 3 we present the new simulation pipeline and in
Sect. 4 how MOAO PSFs are simulated; In Sect. 5 we describe the
kinematic measurements; In Sect. 6 and 7, we present the simulations
and their results, which are discussed in Sect. 8. In Sect. 9, we draw
the conclusions of this study.

\section{Specifying MOAO-fed spectrographs}

\subsection{MOAO \& 3D spectroscopy: EAGLE and FALCON}

Most of current dynamical studies of distant galaxies are severely
hampered by their limited spatial resolution, which introduces an
important uncertainty as to their dynamical nature as discussed in the
Introduction. A MOAO-fed integral field spectrograph on the VLT and/or
the E-ELT will improve substantially the spatial resolution compared
to non-AO assisted observations. The concept of MOAO was first
proposed in the FALCON design study, a multi objects 3D spectrograph
intended for the VLT \citep{Hammer02}. Briefly, the main objectives of
the FALCON concept was to obtain 3D spectroscopy of distant emission
line galaxies, up to $z \sim 2$. Doing this requires us to observe in
the near infrared since all of the important optical emission lines
from [OII]$\lambda$3727\AA~ to H$\alpha$ are redshifted into the
atmospheric windows from 0.8 to 2.5 $\mu$m (see, e.g.,
\citealt{Hammer04}). Because the intrinsic size of distant galaxies is
small, AO is needed to spatially resolve emission lines and thus to
keep the spectroscopic signal-to-noise ratio (SNR) high enough to
detect small features. Correcting a large field of view (FoV) to
observe many galaxies simultaneously is challenging and normal single
natural guide star AO is not sufficient. MOAO is a method for
obtaining excellent correction over a relatively large FoV ($\geq$few
arcmin). The basis of this concept is to correct only regions where
the targets lie, instead of correcting the entire FoV of the
instrument. MOAO uses atmospheric tomography techniques
\citep{Tallon90,Tokovinin01} meaning several wave-front sensors (WFS)
for each integral field unit (IFU) measure the off-axis wavefront
coming from stars located within the FoV, and the on-axis wavefront
from the galaxy is deduced from off-axis measurements and corrected by
an AO system within the optical train of each IFU. A detailed
description of FALCON can be found in \cite{Assemat07}. The general
concept of MOAO was subsequently adapted to the future European
Extremely Large Telescope (E-ELT) in the WFSPEC study
\citep{Moretto06}, which was a precursor to the current EAGLE concept.
One of the most important goals in setting the science design
requirements for EAGLE is mapping the physical properties of galaxies
up to z$\sim$5, in order to better understand the processes that drive
the evolution of galaxies.

\subsection{Goals of this study}

As a first study, we want to help derive high-level design specifications
for the EAGLE and FALCON MOAO-fed instruments by addressing the following:

\begin{itemize}
\item Explore the necessary order of magnitude of integration times
that will be required to reliably derive the kinematics of distant
emission line galaxies, allowing their classification;

\item By selecting a simple rotating disk with little substructure
and a merger simulation with two overlapping disks, explore the most
relaxed requirements for the design of the instrument in terms of the
necessary adaptive optics performance and coarseness of the pixel scale.
To this end we have limited our simulations to spatial sampling of 50-75
mas pixel$^{-1}$;

\item Help constrain the requirements on the MOAO system by specifying
its spectroscopic coupling (which is defined here as the Ensquared
Energy [EE] per element of the spatial resolution, see Sect. 4).
\end{itemize}

A detailed study of the scientific capabilities of these instruments
is beyond the scope of this paper. Instead, we want to use a limited
number of comparisons between scientifically motivated cases in order
to help derive the most relaxed high-level specifications.

\subsection{Methodology}
Flux is the zero-order moment of an emission line, while the velocity and
width are the first and second moment, and higher order moments always
have more relative uncertainty.  For this reason, kinematics will set
stringent requirements on the SNR necessary to fully characterize faint
extended emission line gas.

\begin{figure*}%[h!]
\centering
\includegraphics[width=16cm]{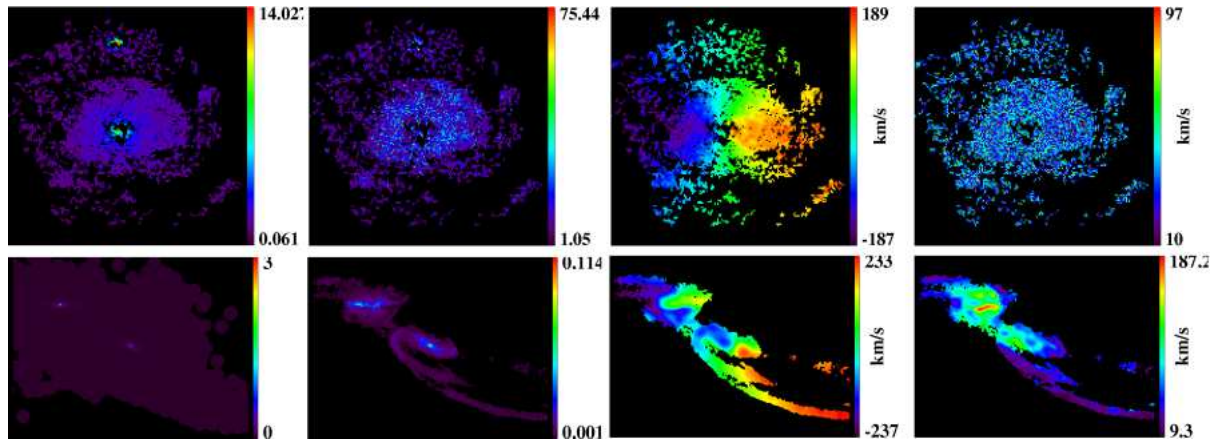}
\caption{Morphology and kinematics of the two objects used in the
simulations. \emph{Up:} UGC5253 (rotating disk). \emph{Down:} Merging
between two Sbc galaxies. \emph{From left to right:} continuum map,
monochromatic map, velocity field, and velocity dispersion map. The
original spatial scale is $\sim$85 pc/pix for UGC5253, and $\sim$200
pc/pix for the merger simulation. These templates are rescaled in
terms of size and flux to simulate realistic distant galaxies.}
\label{local}
\end{figure*}

The characteristics of these moments depend on the underlying source
of an emission line's excitation and kinematics.  Because the most
easily reached goal of MOAO-fed spectrographs will be to understand the
kinematics of the emission line gas in galaxies, a natural kinematic
reference case for such a study is a simple spiral rotating disk. In
the following we will use Fabry-Perot observations of the local rotating
disk UGC5253 taken as part of the GHASP survey (\citealt{Amram02}; see
its main photometric and dynamical properties in Tab. \ref{TabGHASP};
see also Fig. \ref{local}) as such a reference case.

\begin{table}
\centering
\begin{tabular}{ccccccc}\hline\hline
UGC & Type & z & D(') & M$_B$ & inc($\deg$) & V$_{max}$\\\hline
5253 & SA(rs)ab & 0.004306 & 4.168 & -20.8 & 40 & 261.8\\\hline\hline
\end{tabular}
\caption{Main photometric and dynamical properties of UGC5253, used as
a typical rotating disk in the simulations: Morphological type (from
\citealt{Garrido02}), redshift (from CDS), absolute B-band magnitude,
inclination, and maximal velocity of rotation (from
\citealt{Garrido02}).}
\label{TabGHASP}
\end{table}

Several different mechanisms have been proposed to explain how
galaxies assembled their mass: major mergers (merging between galaxies
having similar masses), minor mergers (merging between galaxies with a
mass ratio larger than 5:1), and gas accretion (of cold or hot gas
from the IGM or dark matter halo). For each of these processes, there
are a large number of possible configurations or characteristics that
need to be modeled or analyzed (e.g., mass ratio of merging galaxies,
the relative angular momentum vectors of each merging galaxy and of
the orbit, the temperature and density distribution of the gas as it
is being accreted, etc.). However, our goal here is not to conduct a
systematic study of the scientific capabilities of MOAO-fed
spectrograph in discriminating between all possible evolutionary
mechanisms. Instead, we want to investigate the surface brightness
limits of such instruments and focus on the possibility to use a
relatively coarse spatial resolution to recover large scale motions,
whereby the process underlying the dynamics may be divined (following
for example \citealt{Flores06, Puech06a, Yang07}). Because we only
want to help determine plausible design requirements relative to this
goal, it is sufficient to explore the most extreme large-scale
kinematic patterns, which are produced by major mergers. Indeed,
during such processes, the disks of the progenitors are usually
completely destroyed, which leads to complex kinematic patterns with
abrupt changes in the velocity gradient(s). In the following, we will
use a major merger extracted from an hydrodynamic simulation of a pair
of two merging Sbc galaxies \citep{Cox06}, just before the bodies of
the two galaxies coalesce, but are still discrete entities, which also
provides us with the relaxed requirements in terms of spatial
resolution (see Fig. \ref{local}). In addition, the relatively low
surface brightness extended gas produced during such events (see,
e.g., the tidal tail in Fig. \ref{local}) will also allow us to
explore the ability of MOAO in recovering such faint discrete
features. Trying to detect these features again emphasizes our choice
of simulations and is really leading toward constraints that optimize
for surface brightness detection and away from resolving structures
within disks and in the chaotic central regions of advanced mergers.
In the next three sections, we describe the different steps of this
study: (a) Simulation of data-cubes (Sect. 3); (b) Simulations of MOAO
PSFs (Sect. 4); (c) Kinematic analysis of the data-cubes (Sect. 5).

\section{An end-to-end simulation pipeline for 3D spectroscopy}

\subsection{General Description}

The main steps of the process can be summarized as follows. First, a
data-cube with with the spatial resolution of the telescope
diffraction limit (i.e., $\sim \lambda/2D$, where D is the telescope
diameter) is generated. For each pixel of this high resolution
data-cube, a spectrum is constructed from observations of local
galaxies or results from numerical simulations. In the second step the
spatial resolution of the data-cube is reduced by convolving each
spectral and spatial pixel of the high resolution data-cube by a PSF.
This PSF used in this convolution is representative of the optical
path through the atmosphere and the telescope up to the output of the
(optional) AO system. In the third step, the spatial sampling of the
data-cube is reduced to match that of the IFU of the simulated
instrument. Finally, realistic sky as well as photon and detector
noise are added. Note that this software can also be used to simulate
images of distant galaxies, since an image can be viewed as a
particular case of a data-cube, with a single broad-band spectral
channel. We now describe each step in detail; all input parameters
are summarized in Table \ref{TabParam}.

\subsection{High resolution data-cube}

At very high spatial resolution, emission lines with kinematics driven
only by gravitational motions is well described by a simple Gaussian
\citep{Beauvais99}. Under this assumption, for each pixel of the
data-cube, only four parameters are required to fully define a spectral
line. The first three parameters are the position in wavelength, the
width, and the area (or, equivalently the height) of the emission
line. The current version of the software only models emission lines
and does not take into account the detailed shape of the continuum in
galaxies -- it is simply modeled as a constant in f$_{\lambda}$. So
only one parameter is required to set the level of this
pseudo-continuum around the emission line. We assumed a perfect
Atmospheric Dispersion Corrector (ADC) and did not take into account
atmospheric refraction effects (see, e.g., \citealt{Goncharov07}).
Each spectrum is generated in the observational frame, at a given
spectral sampling of $\lambda _{obs}$/$2R$, where $R$ is the
spectroscopic power of resolution of the instrument, and $\lambda
_{obs} \sim (1+z) \lambda _{em}$, $\lambda _{em}$ being the rest-frame
wavelength of the emission line, and $z$ the redshift of the simulated
object. During this process, the rest-frame line width is multiplied
by (1+z), as one needs to take into account the widening of emission
lines with increasing redshift.

All four parameters can be extracted from observations of local
galaxies. In the following, we use Fabry-Perot (FP) observations of
the H$\alpha$ emission distributions of nearby galaxies obtained as
part of the GHASP survey \citep{Amram02}. From these data, we can
extract four parameters, wavelength, width, area, and continuum level
to construct the velocity field, the velocity dispersion map, the flux
map of the H$\alpha$ emitting gas, and the continuum map of the
galaxy. The software first re-scales all these maps at a given angular
size (in arcsec) provided by the user. These maps are then
interpolated at a spatial sampling of $\sim \lambda/2D$. This spatial
sampling is motivated by the fact that the AO PSFs used have been
simulated at this sampling (see \S4). The software then re-scales the
overall amplitude of the continuum map at a given integrated number of
photons using an integrated magnitude $m_{AB}$ directly provided by
the user. This magnitude is converted to the number of photons per
spectral pixel depending on the telescope diameter $D$, the
integration time $t_{intg}$, and the global transmission $t_{trans}$
of the system (atmosphere excluded). The overall amplitude of the
H$\alpha$ map is also rescaled with a given integrated number of
photons, derived from the integrated continuum value and a rest-frame
equivalent width $EW_0$ provided by the user, the latter being
re-scaled in the observed-frame by multiplying by (1+z).

All parameters can also be extracted from outputs of hydro-dynamical
simulations \citep{Cox04}. In this case, the last two parameters (area
and continuum level) can respectively be extracted from total gas and
stellar surface density maps which are by-products of the
numerical simulation, also rescaled in terms of size and flux.

\subsection{Modeling the IFU and the detector}

Each monochromatic slice of the high resolution data-cube is convolved
by a PSF with matching spatial sampling. This PSF must be
representative of all elements along the optical path, from the
atmosphere to the output of an (optional) Adaptive Optic system.
Because the isoplanetic patch (\citealt{Fried81}; the median value at
Paranal is $\sim$2.4 arcsec at $\lambda \sim$ 0.5 $\mu$m, which leads,
e.g., to $\sim$10 arcsec at $\lambda \sim$ 1.6 $\mu$m) is larger than
the individual FoV of the IFU (typically a few arcsec, depending on
the size of objects at a given redshift), the same PSF can be used to
convolve the data-cube regardless of position within the IFU. We also
neglected the variation of the PSF with the wavelength (with a FWHM
varying as $\lambda ^{-1/5}$ in a Kolmogorov model of the atmospheric
turbulence, see, e.g., \citealt{Roddier81}), as we are only interested
in the narrow spectral range around a single emission line.

The next step is to reduce the spatial sampling of the data-cube. This
is done by re-binning each monochromatic channel of the data-cube at
the pixel size $\Delta _{pix}$ of the simulated IFU. A wavelength
dependent atmospheric absorption curve taken from ESO
Paranal\footnote{www.eso.org/observing/etc} is then applied to each
spectrum of the data-cube. Sky continuum, detector dark level and bias
are then added to the spectra. We used a sky spectrum model (including
zodiacal emission, thermal emission from the atmosphere, and an
average amount of moonlight) from Mauna
Kea\footnote{www.gemini.edu/sciops/ObsProcess.obsConstraints/ocSkyBackground.html},
which has the advantage over other available sky spectra to be very
well sampled with 0.2 \AA/pixel. Photon and detector noise are then
added to each individual exposure. The detector noise is due, in the
NIR, to the dark current ($dark$), and readout noise ($ron$).
%, and the Charge Transfer Efficiency (CTE), which
%is important for optical arrays. CTE is modeled as $\sigma _{CTE} =
%\sqrt{2(1-CTE^p)*N_{photons}}$, where $N_{photons}$ is the number of
%photons in a given spectral pixel, and $p$ is one fourth of the
%detector size (in pixels).
The simulation pipeline generates $ndit$ data-cubes with individual
exposure time of $dit$, which are combined by estimating the median of
each pixel to simulate several individual realistic exposures. Since
we have only included random noise, it is similar to having dithered
all of individual exposures and combining them after aligning them
spatially and spectrally. The spectroscopic SNR is then derived as
follows:

$$
SNR(i_x,j_y,k_{\lambda})=\frac{O(i_x,j_y,k_{\lambda})*\sqrt{ndit}}{\sqrt{O(i_x,j_y,k_{\lambda})+S(i_x,j_y,k_{\lambda})+ron^2+dark}},
$$ where $O(i_x,j_y,k_{\lambda})$ and $S(i_x,j_y,\lambda)$ are
respectively the object and sky flux per $dit$ (after accounting for
atmospheric transmission) in the spatial position $(i_x,j_y)$ of the
data-cube (in pixels), and at the spectral position $k_{\lambda}$
along the wavelength axis (in pixels). In the following, the ``maximal
SNR in the emission line in the pixel $(i_x,j_y)$'' refers to
MAX$_{k_{\lambda}}$[$SNR(i_x,j_y,k_{\lambda})$], and the ``total SNR''
refers to the flux weighted average of this quantity over the galaxy
(i.e., the spatially integrated SNR in the spectral pixel where the
emission line peaks, and \emph{not} the spatially integrated SNR
within the whole emission line width, as sometimes used in other
studies).

\begin{table*}
\centering
\begin{tabular}{lccc}\hline\hline
Parameter & Description (Unit) & EAGLE/E-ELT & FALCON/VLT\\\hline
$M1$ & Telescope Primary mirror size (m) & 42 & 8.2\\
$M2$ & Telescope Secondary mirror size (m) & 0 & 1.116\\\hline
$R$ & Spectral resolution power & 5000 & 5000\\
$\Delta _{pix}$ & IFU pixel size (arcsec) & 0.075'' & 0.125''\\
$t_{transm}$ & Telescope + Instrument transmission factor & 0.2 & 0.2\\\hline
%$CTE$ & Charge Transfer Efficiency & 1 & 1\\
%$p$ & Number of pixels used representative of the CTE (size/fraction of detector) &  4000x4000/4 & 4000x4000/4\\
$dark$ & Dark level (e/sec/pix) & 0.01 & 0.01\\
$ron$ & Read-out-noise (e/pix) & 2.3 & 2.3 \\\hline
$dit$ & detector integration time (s) & 3600 & 3600\\
$ndit$ & Number of dit per simulation & 200-24-8 & 200-24-8\\\hline
$z$ & Redshift of the object & 4.0 & 1.6\\
$S$ & Object angular diameter (arcsec) & 0.8& 2.0\\
$m_{AB}$ & Integrated magnitude of the object continuum (AB mag) & 24.5 & 22.5\\
$\lambda _{em}$ & Emission line wavelength (\AA) & 3727& 6563\\\
$EW_0$ & Equivalent width at rest of the emission line (\AA) & 30 & 50\\\hline\hline
\end{tabular}
\caption{\emph{First and second columns:} Input parameters of the
simulation pipeline provided by the user. \emph{Third column:}
parameters used for the simulations of z=4 galaxies with EAGLE on the
E-ELT. \emph{Fourth column:} parameters used for the simulations of
z=1.6 galaxies with FALCON on the VLT.}
\label{TabParam}
\end{table*}

\section{Generating MOAO PSFs}

The coupling between the MOAO system and the 3D spectrograph is
captured through the MOAO system PSF. Therefore, it is a crucial
element that needs to be carefully simulated, and cannot be
approximated by, e.g., a simple Gaussian. We describe here how these
PSFs were modelled, and briefly describe how the MOAO correction
influences the PSF shape at the spatial scale of the instrument IFU.

\subsection{MOAO PSFs for EAGLE}

\begin{figure}
\centering
\includegraphics[width=8cm]{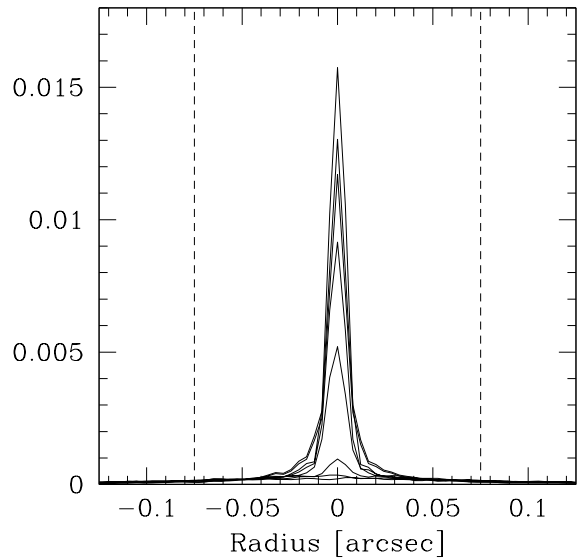}\\
\caption{Transverse cut of the simulated EAGLE MOAO PSFs. The EE in
  0.15 arcsec increases from bottom to top: 22\%, 24\%, 26\%, 31\%,
  34\%, 40\%, 43\%, and 47\%. The vertical dashed-lines represent the
  spatial element of resolution used for EAGLE simulations, i.e., 0.15
  arcsec.}
\label{PSF_elt}
\end{figure}

A preliminary study of the PSF shape generated by a hypothetical MOAO
system can be found in \cite{Neichel06}. In the following, we used
eight PSFs generated in a similar manner as in the Neichel et al.
study. Briefly, three off axis guide stars, located at the edges of an
equilateral triangle, are used to perform a tomographic measurement of
the turbulent atmospheric volume. We have used a turbulent profile
typical of that on Cerro Paranal, which can be modeled using three
equivalent layers \citep{Fusco99}. The seeing and the outer scale of
the turbulence were respectively set to 0.95 arcsec (at 0.5 $\mu m$,
the $\sim$0.75 percentile on Paranal) and to 22m (the median value on
Paranal). Guide stars are assumed to be bright (V$\sim$13), i.e., WFS
noise is neglected. The optimal correction is deduced from the
characteristics of the turbulence volume and applied assuming a single
Deformable Mirror (DM) per direction of interest, here taken as the
center of the guide star constellation. To explore a wide range of
correction, we consider different GS-constellation sizes, as well as a
DM with an inter-actuator size of either 1 m or 0.75 m in the pupil
plane (for a 42 meter telescope), which corresponds to 42x42 or 56x56
actuators (see Tab. \ref{PSFparamELT}). Of note, the SPHERE eXtreme AO
(XAO) project \citep{Beuzit06} will use a 41x41 actuators DM
\citep{Fusco06}. PSFs are shown in Figure \ref{PSF_elt}.

\begin{table}
\centering
\begin{tabular}{cccc}\hline\hline
Pitch & FoV$_{WFS}$ & EE & Strehl Ratio (\%)\\\hline
1.00 & 4.00 & 22 & 0.1\\
1.00 & 3.00 & 24 & 0.2\\
1.00 & 2.00 & 26 & 0.6\\
1.00 & 1.00 & 31 & 3.9\\
1.00 & 0.50 & 34 & 6.7\\
1.00 & 0.25 & 40 & 11.5\\
1.00 & 0.00 & 43 & 10.5\\
0.75 & 0.00 & 47 & 11.7\\\hline\hline
\end{tabular}
\caption{AO system parameters used for the simulation of the EAGLE
PSFs. The EE is given in 0.15 arcsec, and the pitch corresponds to the
inter-actuator distance on the telescope pupil, in meters. The second
column gives the diameter of the constellation of three stars (at
equal distance of the central galaxy) used to sense the wavefront, in
arcmin. Strehl Ratios are given in the last column.}
\label{PSFparamELT}
\end{table}

Usually, AO PSFs are characterized using the Strehl Ratio [SR], which
provides a useful way for comparing different PSFs relatively to the
diffraction limited case. However, MOAO does not provide diffraction
limited PSFs but performs only partial corrections in order to
increase the spectroscopic coupling with the 3D spectrograph
\citep{Assemat07}, which cannot be easily described by using the SR.
Instead, the spectroscopic coupling can be directly quantified by the
fraction of light under the PSF that enters a spatial element of
resolution of the IFU, i.e., the Ensquared Energy entering an aperture
equal to twice the IFU pixel size. Such a parameter is an
extrapolation of the classical ``Encircled Energy'' used in optics to
characterize optical system quality, and is directly related to the
achieved SNR. It is therefore a more natural choice to parametrize the
MOAO performance than the SR. In the presence of speckle noise on the
PSF as it is the case in the end-to-end MOAO simulations used here,
the SR measurement can be severally affected by this noise, which
produces artificially low SR values and contributes to disconnect the
SR measurement from the spectroscopic coupling: at a given EE, a PSF
affected by speckle noise will have a lower SR than a PSF without
speckle noise, all else being equal. We list both SR and EE for the
simulated PSFs in Tab. \ref{PSFparamELT}.

\begin{figure}
\centering
\includegraphics[width=8cm]{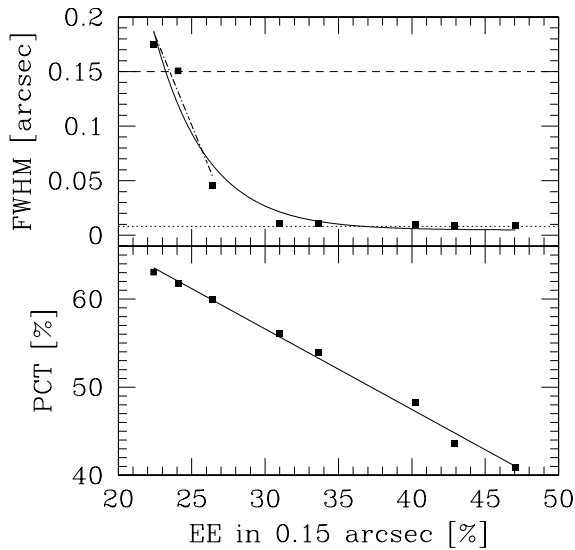}
\caption{\emph{Top:} FWHM vs. EE (in 0.15 arcsec) for the eight
simulated EAGLE PSFs. The dotted line represents the diffraction limit on
a 42m ELT in H band, i.e., $\lambda _{diff} (H)=0.008$ arcsec. The
black line is an exponential fit between the FWHM and the EE, and the
dotted-dashed line is a simple linear fit to the three first points. The
dashed lines shows when the FWHM becomes lower than 0.15 arcsec, i.e.,
twice the pixel size (under-sampling of the PSF). \emph{Bottom:} PCT
vs. EE for the eight simulated EAGLE PSFs. The black line is a simple
linear fit between PCT et EE.}
\label{EE_fwhm_elt}
\end{figure}

To explore how the AO correction impacts the PSF shape, we first
derived the PSF FWHM using Sextractor \citep{Bertin96}. We compared
these to their Ensquared Energy [EE] in a aperture equal to twice the
pixel size (i.e., 0.15 arcsec) in Fig. \ref{EE_fwhm_elt}. The EE is
measured as the fraction of light under the PSF that enters this
aperture. Fig. \ref{EE_fwhm_elt} reveals two regimes, depending on the
EE. In the first regime, the EE and the FWHM are roughly linearly
correlated, so that the EE can alternatively be used as an indirect
measure of the FWHM. This regime corresponds to the formation of a
central diffraction limited core in the PSF. When the adaptive system
does not correct for the overall atmospheric turbulence, it is indeed
well known that the resulting PSF can be approximately described by a
double core-halo structure, with a coherent central
diffraction-limited peak surrounded by seeing-limited wings. This
regime corresponds to an improvement of the spatial resolution, as
long as the PSF FWHM remains larger than twice the IFU pixel scale.
When the PSF FWHM becomes smaller than twice the IFU pixel size, the
spatial resolution is now set up by the IFU pixel scale. In this case,
the AO corrections does not dramatically influence the spatial
resolution any longer, but can still provide better EE and SR by
further narrowing the core below the IFU pixel scale (see Tab.
\ref{PSFparamELT}), which translates into an improvement of the SNR.
When EE$\geq$27-30\%, the diffraction limited core is now well formed,
which explains why the measurement of the FWHM of the PSF saturates,
since the FWHM is then only sensitive to the PSF central core as it
does not ``see'' the residual halo around it. In this new regime, the
SR keeps rising with the EE because the AO system is still able to
bring energy located in the PSF halo into the central diffraction
limited peak. In most situations, the MOAO system will provide EE
larger than this limit, which justifies the choice of characterizing
the EE in an aperture equal to twice the pixel size.

In the intermediate regime where the PSF FWHM is smaller than twice
the IFU pixel scale, but where the diffraction core is not completely
formed (e.g., EE ranging between 23 and 35\% in Fig.
\ref{EE_fwhm_elt}), the PSF halo can be an important limitation in the
observations of high redshift galaxies, because it can result in a
mixing of the spectra coming from adjacent spatial elements of
resolution. In the following, we use the total energy entering the
first ring of 0.15 arcsec apertures around the one centered on the PSF
to measure this effect. This provides a useful proxy to quantify the
amount of light coming from adjacent spatial element of resolution,
and polluting the element of resolution centered on the PSF. We called
this effect ``PSF cross-talk'' (PCT), by analogy with the instrumental
crosstalk, which measures the amount of scattering that spreads
information beyond the PSF due to instrumental effects. Figure
\ref{EE_fwhm_elt} also compares the PCT with the EE for the set of
eight simulated PSFs, and shows a linear correlation between these two
parameters: while the AO system provides a better correction, the
energy located in the wings of the PSF is brought into the central
diffraction-limited core (see \citealt{Assemat07}), increasing the EE
and decreasing the PCT. PCT gives as useful measurement of the ``PSF
contrast'': having a PCT smaller than 50\% means that the light
entering a spatial element of resolution is not dominated by polluting
light coming from adjacent spatial elements of resolution. In the
EAGLE case and at a pixel scale of 75 mas, this occurs at EE$\sim$35\%
(see Fig. \ref{EE_fwhm_elt}).

\subsection{MOAO for FALCON}

\begin{figure}
\centering
\includegraphics[width=8cm]{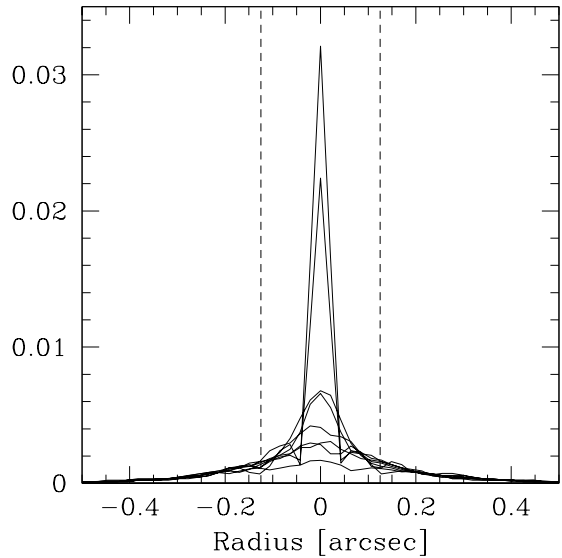}\\
\vspace{0.2cm}
\caption{Transverse cut of the FALCON MOAO PSFs simulated. The EE in
  0.25 arcsec increases from bottom to top: 19\%, 26\%, 27\%, 31\%,
  34\%, 34\%, 35\%, 39\%, and 46\%. The vertical dashed-lines
  represent the spatial element of resolution used for FALCON
  simulations (0.25 arcsec aperture) within which EE are derived.}
\label{PSF_falcon}
\end{figure}

An extensive study of the FALCON MOAO system has been carried out by
\cite{Assemat07}. As part of this study, a simulation pipeline of
FALCON PSF was developed. We used this tool to generate a set of eight
PSFs (see Fig. \ref{PSF_falcon}). All PSFs were simulated using
typical atmospheric conditions at ESO/Paranal. Briefly, a median
seeing of 0.81 was used, with an outer scale of the turbulence of 24m,
and an atmosphere model composed of three layers located at [0,1,10]km
with respective weights of [20, 60, 20]\%. We set the parameters of
the MOAO system in order to explore a wide range of EE in 0.25 arcsec
(twice the pixel size), ranging from 19\% to 46\% (see Tab.
\ref{PSFparamFALCON}). Technical details about the concept can be
found in, e.g., \cite{Puech05}.

\begin{table}
\centering
\caption{AO system parameters used for the simulation of the FALCON
  PSFs. The EE is given in 0.25 arcsec, and the value quantifying the
  correction corresponds to the radial number n of Zernike modes
  corrected. The third column gives the diameter of the constellation
  of three stars (at equal distance of the central galaxy) used to
  sense the wavefront, in arcsec. Following Puech et al. (2005), n=7
  corresponds roughly to a $\sim$1.4m pitch, and n=9 to a $\sim$1.1m
  pitch. The first PSF is the seeing limited case, with no correction.
  Strehl Ratio are given in the last column.}
\begin{tabular}{cccc}\hline\hline
Correction & FoV$_{WFS}$ & EE & Strehl Ratio (\%)\\\hline
0 & N.A. & 19 & 1.03\\
7 & 300  & 26 & 1.56\\
7 & 180  & 27 & 1.51\\
7 & 60   & 31 & 2.26\\
7 & 30   & 34 & 4.04\\
7 & 10   & 35 & 4.55\\
7 & 60   & 39 & 20.82\\
9 & 60   & 46 & 30.03\\\hline\hline
\end{tabular}
\label{PSFparamFALCON}
\end{table}

\begin{figure}
\centering
\includegraphics[width=8cm]{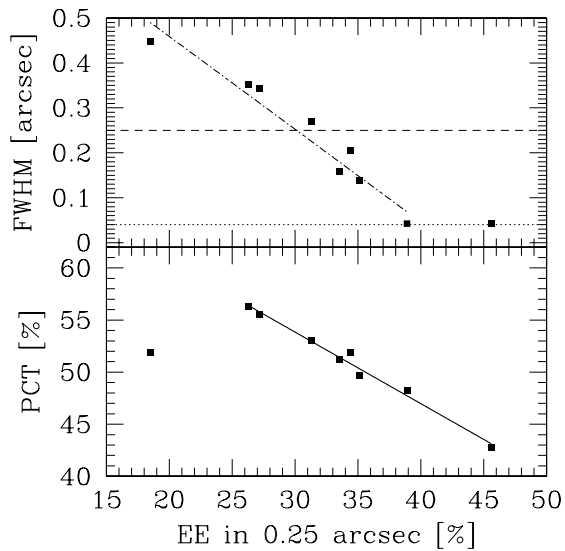}
\caption{\emph{Top:} FWHM vs. EE for the eight simulated FALCON PSFs.
The left most point corresponds to the seeing limited PSF. The
horizontal line represents the diffraction limit on the VLT in H band,
i.e., $\lambda _{diff} (H)=0.04$ arcsec. The dashed lines show when the
spatial resolution, as measured by the FWHM, is lower than 0.25
arcsec, i.e., twice the pixel size. The black line is a linear fit
between FWHM and EE (the last point, where the FWHM saturates, has
been discarded from the fit). \emph{Bottom:} PCT vs. EE for the eight
simulated FALCON PSFs. The left most point corresponds to the seeing
limited PSF. The black line is a linear fit between the PCT and the
EE.}
\label{EE_fwhm}
\end{figure}

For this set of PSFs, we compared their FWHM and PCT to their EE in
0.25 arcsec in Figure \ref{EE_fwhm}, which reveals a similar behavior
compared to Fig. \ref{EE_fwhm_elt}. Note that when EE$\geq$30\%, the
FWHM becomes lower than 0.25 arcsec (i.e., twice the pixel size). This
means that when EE$\geq$30\%, the spatial resolution is oversampled at
twice the IFU pixel size. This is why, in the following, all EE are
computed in a 0.25 arcsec size aperture. Finally, PCT becomes smaller
than 50\% for EE$\sim$35\%, which is the limit for a good ``PSF
contrast'' in the FALCON case, at a pixel scale of 0.25 arcsec.

\section{Simulating and measuring 2D kinematics}

\subsection{Kinematic measurements}
%In a first step, fake observations of distant galaxies are produced
%using high resolution templates of real local observations or from
%hydrodynamical simulations. Concretely, these fake observations result in simulated data-cubes produced as described in Sect. 3. 
The simulated data-cubes are analyzed using a data analysis pipeline
similar to those generally used to analyze data of high redshift
galaxies. During this process, each spatial pixel of the simulated
data-cube is fitted with a Gaussian in wavelength, whose position and
width correspond to the velocity and velocity dispersion of the gas in
this spatial pixel. Because the accuracy of these measurements is
driven by the flux within the emission line relative to the noise in
the continuum, one has to define a kinematic signal-to-noise ratio
$SNR_{kin}$ specific for this measurement. We chose to rely on the
definition of \cite{Flores06}, who defined $SNR_{kin}$ as the total
flux in the emission line divided by the noise on the continuum
$\sigma _{continuum}$ and $\sqrt N_{pix}$, the number of pixels within
the emission line. We emphasize that this $SNR_{kin}$ is different
from the classical $SNR$, as defined in Sect. 3.3: the latter
quantifies the signal to noise ratio relative to the sky and
detector noise (detectability of the emission line), while the former
characterizes the capability of measuring the first and second order
moments of the emission line, relative to the noise in the continuum
\citep{Sarzi06}. Although they are different in nature, we find that
both SNRs roughly correlates provided that $SNR>3$ (corresponding the
$SNR_{kin} \sim$3-4). Because a lot of data-cubes are simulated and
need to be analyzed, one has to rely on a fully automatic analysis
pipeline. Tests show that pixels with a ``kinematic'' $SNR_{kin} <$ 3
have very large errors mostly due to misidentification of emission
lines with noise peaks. Hence, only pixels having a $SNR_{kin}$ of at
least three have been kept for further analysis.

\subsection{Kinematic diagnostic diagrams}
To help distinguish between the two kinematic templates (i.e., the
major merger and the rotating disk), we fitted a rotating disk model
to each velocity field. This model assumes a constant dynamical
center, PA, inclination, and $V_{sys}$ over the galaxy, and a simple
\emph{arctan} shape for the rotation curve. During the fit, the pixels
of the velocity field are weighted by the corresponding error in bins
of SNR, which were determined using Monte-Carlo simulations. Finally,
the residuals between the velocity field and the fitted model are
derived, together with the reduced $\chi ^2$ map.

We then adopted two criteria for judging whether or not the fit to a
particular simulated galaxy is appropriate.
%In a given pixel,
%a normalised residual of the fit to each pixel value smaller than
%unity means that, in this pixel, the model is acceptable within the
%uncertainty. 
Because we are interested only in large-scale motions (i.e., circular
rotation vs. non-structured motions), we need to adopt a criterion not
affected by smaller-scale variations due to non-circular perturbations
like, e.g., spiral arms or bars. For characterizing the 2D-fit as a
whole, we therefore rely on the median $\chi ^2_m(VF)$ over the
galaxy. When smaller than unity, it means than at least 50\% of the
pixels on the velocity field are reliably fitted by a rotating disk
model, and that galaxy motions are actually dominated by circular
rotation. In addition, we choose to compare the distance between the
fitted dynamical center and the velocity dispersion barycenter. We
used such a criterion because in the case of data which are heavily
smoothed by the seeing, a process that leads to more complex dynamics
like merging would only rarely have its dynamical center of the entire
system coincident with the maximum in the distribution of velocity
dispersions. For this, we derived the associated uncertainty, using
classical propagation of errors due to the uncertainties on the
dynamical center fit and on the sigma barycenter. We then derived a
parameter $\Delta \sigma _{peaks}$ defined as the distance between the
fitted dynamical center and the velocity dispersion barycenter,
normalized by the corresponding uncertainty. Having $\Delta \sigma
_{peaks} < 1$ means that the fitted dynamical center is found close
enough to the sigma barycenter location relative to the associated
uncertainty, to be considered as compatible with the dynamics of a
rotating disk.

In summary, the kinematics of a galaxy can be considered as compatible
with that of a rotating disk, provided that it falls on a region of a
``diagnostic diagram'' ($\chi ^2_m(VF)$,$\Delta \sigma _{peaks}$)
where the median normalized residual is smaller than one, and provided
that $\Delta \sigma _{peaks} \leq$ 1. We need to emphasize that this
classification (as any) is not without limitations. As an example, let
us consider a merging pair of galaxies. If both progenitors are
equally distant from the center of mass of the system, the velocity
dispersion map will show two peaks equally distant from this center,
and the resulting barycenter of the sigma map may then be found very
close to the dynamical center of orbital motions between the two
progenitors if the separation of the two is small compared to the
intrinsic resolution of the data and the exact distribution of the
emission line gas. Of course, that is a limitation of most techniques
as it can be difficult to identify close merging pairs at high
redshift. It is important to be aware that there are some
configurations that are difficult to distinguish between. Be that as
it may, we use this approach as a first attempt in quantifying the
rotation in very distant galaxies.

\section{Simulations}

\subsection{Assumptions}
It was decided not to simulate observations in the K-band as the
detectability of lines in the thermal infrared is highly dependent on
the achieved emissivity of the instrument. To generalize these first
simulations and to remove additional free parameters, such as the
telescope and instrumental emissivities and temperatures, we consider
the H band only, assuming no thermal emission from the instrument and
the telescope in the H-band. Our tests show that this assumption leads
to overestimate the SNR in the H-band by no more than $\sim$10\%. This
makes these first simulations less dependent of the telescope design
(e.g., number of mirrors), environmental conditions (site selection),
and instrument characteristics (e.g., number of warm mirrors).
Moreover, working in the H-band will help disentangle spatial effects
from flux limitation arising from the high thermal background in the
K-band.

For this study, we assumed a spectral power of resolution $R$=5000 as
a compromise between our desire to minimize the impact of the OH sky
lines and not wanting to over-resolve the line by a large factor.
Moreover, appropriate targets in the NIR are usually selected as they
have emission lines that fall in regions free of strong OH lines. We
estimated that about one third of the H-band is free from strong sky
background variations (i.e., larger that 10\% of the sky continuum) in
continuous windows of at least 200 km/s\footnote{This obviously
depends on the spectral resolution and sampling of the sky background.
These figures were determined using the Mauna Kea sky model introduced
in Sect. 3.3, which has a spectral resolution of 0.4 $\AA$ and is
sampled at the Nyquist rate. This gives a spectral power of resolution
R$\sim$22500 at $\lambda$=9000$\AA$.}. Therefore, for simplicity, all
simulations have been performed without OH sky lines, accounting for
the sky continuum only. Usually, R=3000 is considered as a strict
minimal spectral resolution to work between OH sky lines. In
comparison, R=5000 will allow a better sky subtraction and limit
kinematic measurement (i.e., velocity and velocity dispersion)
uncertainties. In addition, for simplicity, we assume that the [OII]
emission line is a single line, instead of a doublet. This does not
influence any results presented here, as these are scaled with the
\emph{total} flux, i.e. the flux inside both lines of the doublet
(assuming the low density limit for the ratio of the two lines).

The instrumental parameters of the simulation were set to typical
values used in the IR (see Tab.\ref{TabParam}), relying on a cooled
Rockwell HAWAII-2RG IR array working at $\sim$80K, as described by
\cite{Finger06}. An array of this type is already used in SINFONI, and
arrays with similar characteristics will be implemented in next VLT
generation IR instruments such as HAWK-I, KMOS, or X-SHOOTER. 
%With IR
%arrays, there is no CTE since charge is not transferred in these
%devices. 
Table \ref{TabParam} summarizes the main parameters used in
the simulations.

Previous studies of FALCON have considered a pixel scale of 0.125
arcsec \citep{Puech04,Assemat07}. We will thus adopt this scale in the
following. For EAGLE, we chose to firstly explore a pixel scale of 75
mas. This choice is suggested by recent 2D-kinematical studies of
distant galaxies with GIRAFFE, which show that a relatively coarse
pixel scale can already provide us with enough information to properly
recover large scale motions in distant galaxies, and then discriminate
between simple rotating disks and major mergers (see
\citealt{Flores06,Puech06a}).

\subsection{Simulation of z$\sim$4 galaxies with EAGLE}

We conducted simulations at z=4, where the [OII] emission line falls
in the H band. Both objects have been rescaled in half-light radius
and flux to account for evolution and redshift (see bottom of
Tab.\ref{TabParam}). \cite{Ferguson04} and \cite{Bouwens04} found that
the typical r$_{half}$ of z$\sim$4 galaxies is $\sim$0.2 arcsec in the
UV. Hence, we adopted an optical diameter of 0.8 arcsec for z$\sim$4
galaxies ($R_{opt}\sim 2r_{half}$). The continuum magnitude was chosen
considering \cite{Yoshida06}, who studied z$\sim$4 Lyman Break
Galaxies luminosity functions in the UV. They found a typical UV
magnitude of 24.5. We neglected any color corrections between UV and
blue optical ([OII]$\lambda$3727\AA): as spiral Spectral Energy
Distributions [SEDs] rise from UV to optical, this choice can be
considered as a minimal value for a typical pseudo-continuum for
z$\sim$4 galaxies. We assumed an integrated rest-frame
equivalent-width EW$_0$([OII])=30\AA, extrapolating the median values
found in z$<$1 galaxies by \cite{Hammer97}. This leads to an
integrated emission line flux $\sim$7.5$\times$10$^{-18}$erg/s/cm$^2$
(observed-frame).

\subsection{Simulations of z$\sim$1.6 galaxies with FALCON}
We choose as a rest-frame Equivalent Width $EW_0$ for the H$\alpha$
emission line, the minimal value observed in the sample of
\cite{Erb06} at z$\sim$2. In their sample, the continuum at 1500\AA~
ranges from $m_{AB}$ = 22 to 26. Assuming a U-H color of 3.5, typical
of Sa-Sb galaxies \citep{Mannucci01}, and assuming that this value can
be extrapolated at z$\sim$1.6, this leads us to a minimal magnitude in
the H band of $\sim$22.5. As a cross-check, we looked at the sample of
BM/BX galaxies of \cite{Reddy06}. We found 22 sources having a
redshift between 1.4 and 1.8, for which Reddy et al. give J and K
bands AB magnitudes. Assuming that the H-band magnitude can roughly be
approximated by an average between J and K bands magnitudes, we found
in these 22 sources, a median (mean) H-band magnitude of 22.55
(22.56). We can thus assume with some confidence that m$_{AB}$(H)=22.5
is quite representative of luminous z$\sim$1.6 galaxies. This leads to
an integrated emission line flux
$\sim$4.9$\times$10$^{-17}$erg/s/cm$^2$ (observed-frame). Finally,
following \cite{Ferguson04}, we assumed a typical half-light radius
r$_{half}$ of 0.5 arcsec in UV ($\sim$1500\AA) for z$\sim$1.6
galaxies. A similar value was found by \cite{Dahlen07} at 2800\AA,
with 0.4 arcsec for z=1.75 galaxies with M$_* \geq$10$^{10}$M$_\odot$.
We took 2 arcsec as a linear diameter for z$\sim$1.6 galaxies,
neglecting the k-morphological correction between UV and rest-frame
optical (H$\alpha$). The parameters adopted for the simulations are
listed in Tab. \ref{TabParam}.

\section{Results}

\subsection{The Design Requirements for EAGLE on the E-ELT}

\begin{figure*}
\centering
\includegraphics[width=17cm]{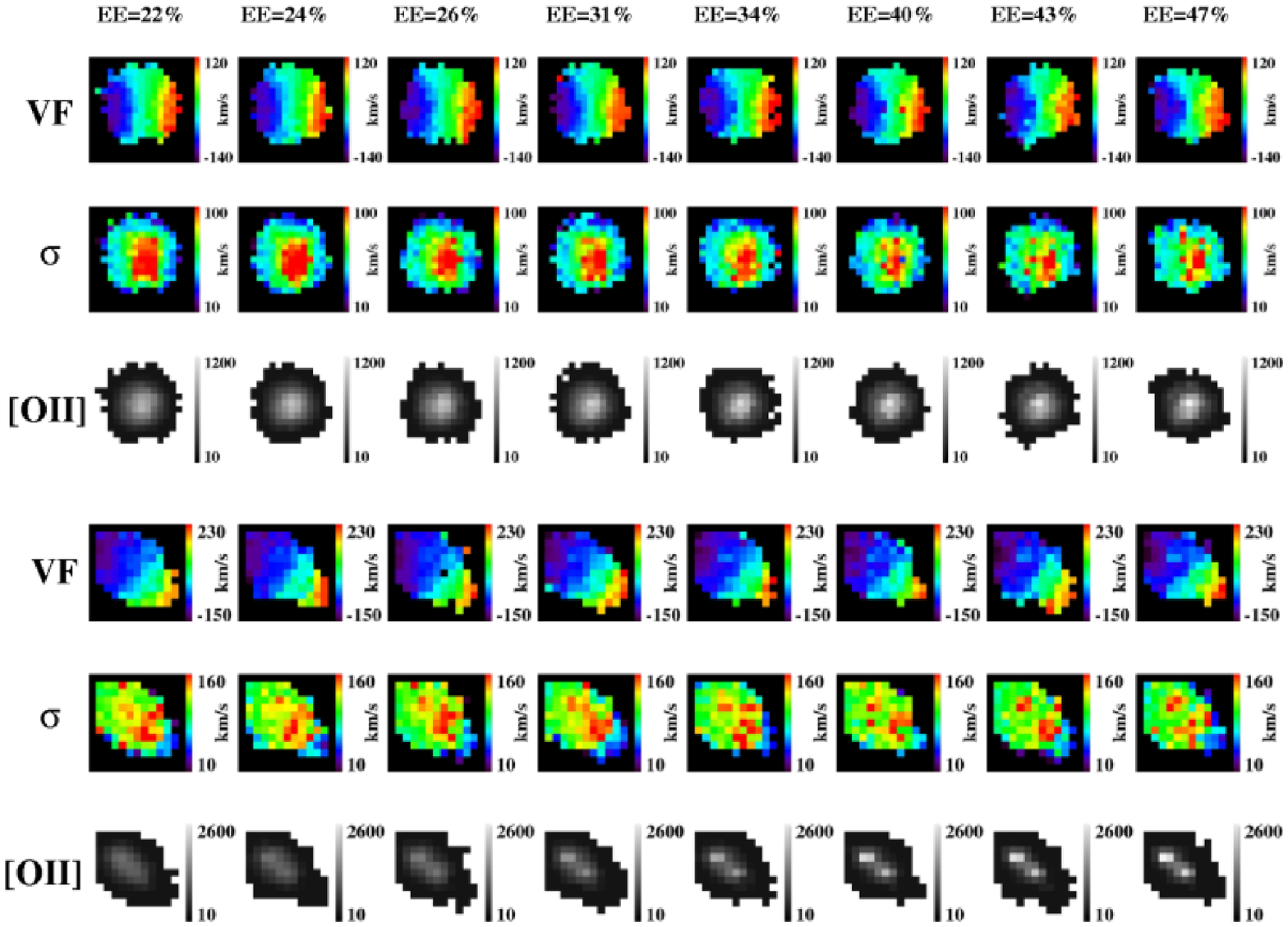}
\caption{Kinematics of a rotating disk (three top lines) and a major
merger (three lowest lines) as derived from simulations of EAGLE
observations at z$\sim$4, with 75 mas pixel$^{-1}$ and
t$_{intg}$=200h. The velocity fields (VF), the velocity dispersion
maps ($\sigma$), and the emission line flux maps ([OII], units are in
median counts per \emph{dit}) are shown. Each column corresponds to a
given EE in 0.15 arcsec, as indicated at the top.}
\label{SimELT200}
\end{figure*}

As explained above, the effects of the PSF on the data-cube are
twofold. On one hand, the FWHM (and the PCT) influences the spatial
resolution of the observations. On the other hand, the EE strongly
influences the number of photons reaching the detector. Because the EE
roughly correlates with the FWHM (and the PCT), both effects are
related. An easy way to distinguish them and to focus on spatial
features is to simulate data-cubes with very large SNRs. We therefore
first ran a set of simulations with t$_{intg} \sim$ 200h: Figure
\ref{SimELT200} shows the velocity fields, velocity dispersion maps,
and emission line flux map for the z$\sim$4 simulated rotating disk.
The effect of the PSF can clearly be seen on each of these three maps
(from left to right). More flux (SNR) if brought toward the galaxy
center, the ``spider'' shape of the isovelocities are more robustly
recovered, and the peak at the dynamical center of the dispersion map
is not so heavily smoothed as the correction increases. These effects
are even clearer in the simulation of the major merger: the two peaks
in the emission flux maps (see also the velocity dispersion maps),
which correspond to the two progenitors, are clearly better
distinguished as the correction improved, as are the non-circular
motions in the velocity field. The two progenitors can be
distinguished provided that EE $>$ 26\%. It is worth to note that this
EE corresponds to a FWHM smaller than 2 pixels (see Fig.
\ref{EE_fwhm_elt}), i.e. to a regime where the central diffraction
peak is already formed (see Sect. 4.1). As the two progenitors are
separated by only 2-3 pixels, a poor correction mixes the two
corresponding peaks in the emission line flux maps and/or the velocity
dispersion maps. Obviously, detecting smaller spatial-scale features
would require a finer pixel scale. However, at a given pixel size,
these simulations suggest that, from a purely spatial point of view
(i.e. at infinite SNR), an optimal way to couple AO with 3D
spectroscopy is to tune the MOAO system in order to get a PSF whose
FWHM is not larger than twice the pixel size. Improving the correction
further will not provide a better spatial resolution, but at that
point a finer pixel scale may help (see Sect. 4). However, even with a
relatively coarse sampling, a higher EE, will in turn lead to a higher
SNR in a given integration time: with realistic integration times, the
optimal choice of AO correction must take into account achieving
sufficient SNR to detect low surface brightness regions.

\begin{figure}
\centering
\includegraphics[width=8cm]{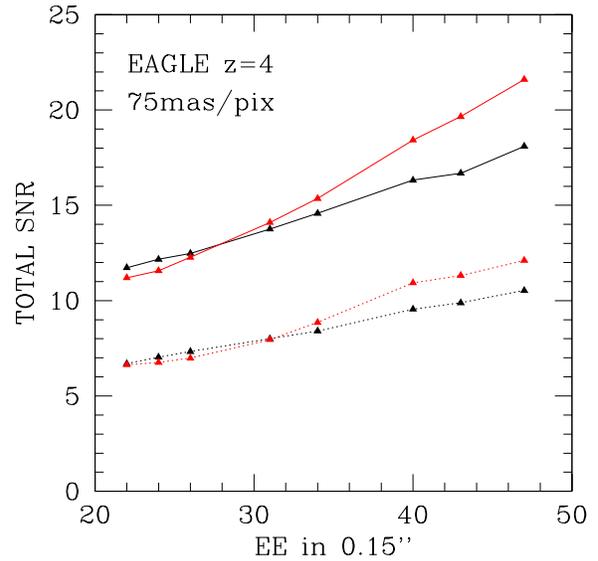}
\caption{Flux-weighted SNR in the center of the emission line,
integrated over the galaxy for EAGLE simulations with 75mas/pix and 24
or 8 hr of exposure times (respectively: full lines and dotted lines).
The black curves correspond to the rotating disk, while the red ones
correspond to the major merger.}
\label{snr_75mas}
\end{figure}

To investigate the impact of the SNR on the AO requirements, we ran
simulations with shorter total integration times: we adopted perhaps
more realistic values of 24 and 8 hours. With an integration time of
24hr, the total SNR over the entire galaxy ranges between $\sim$11 and
$\sim$22, depending on the EE (see Fig. \ref{snr_75mas}). This SNR goes
down between $\sim$7 and $\sim$12 with 8hr of exposure time, scaling
as the square root of the integration time, as expected. For a given
surface brightness distribution and integration time, the SNR scales
linearly with EE, as expected in a background dominated regime. Adopting
a total SNR of 5 as a lower limit to recover reliable and meaningful
information over a galaxy, one can derive that between 1.5 and 4 hours
of integration time is a strict minimum to recover z=4 galaxy kinematics,
depending on the EE provided by the MOAO system and the pixel scale.

\begin{figure}
\centering
\includegraphics[width=8cm]{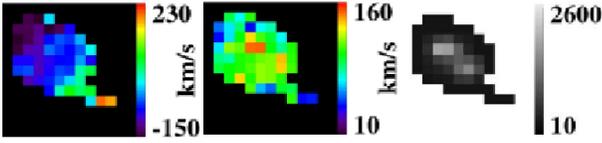}
\caption{Kinematics of a major merger as derived from simulations of
EAGLE observations at z$\sim$4, with 75mas/pixel, t$_{intg}$=24h and
EE=34\%. The velocity fields (first column), the velocity dispersion
maps (second column), and the emission line flux maps (last column,
units are in median counts per \emph{dit}) are shown. The object size
is 0.8 arcsec in diameter, which represents $\sim$0.12 kpc at z=4.}
\label{eagle34}
\end{figure}

\begin{figure}
\centering
\includegraphics[width=8cm]{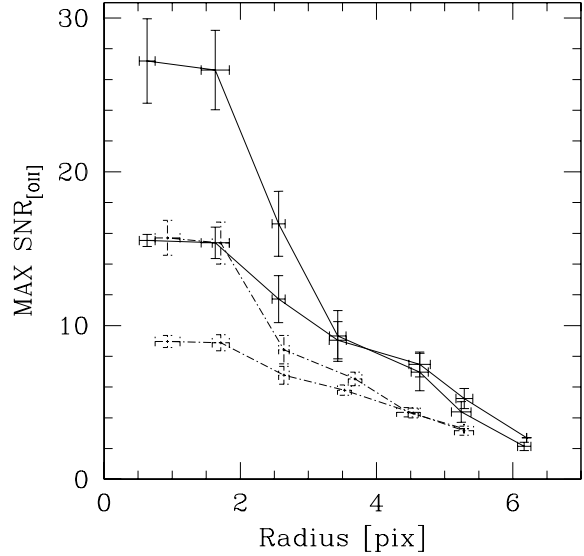}
\caption{Radius-average profile of the maximal SNR in the [OII]
emission line of the rotating disk, obtained after 24hr (full lines)
and 8hr (dotted lines) of exposure time. In each case, the upper curve
is obtained for the highest EE (47\%), and the bottom one for the
smallest EE (22\%). Error-bars account for the 1-sigma dispersion in
SNR and positions within each ring of the profile. Only pixels having
$SNR_{kin}>$3 have been considered to build the profile.}
\label{profsnr_eagle}
\end{figure}

In Fig. \ref{profsnr_eagle}, we plot radial-average profiles of the
maximal SNR reached in the emission line for the rotating disk
simulations, both with 24 and 8hr of exposure time, and with 22 and
47\% of the EE (the smallest and largest values used in the
simulations). This figure shows the effect of increasing the EE for a
given integration time: it mainly increases the SNR in the central
part of the galaxy (which has some structure fine enough to benefit
from good AO correction). On the other hand, the outer and lower
surface brightness regions are much more difficult to recover: with
8hr of exposure time, emission lines are only detected over only a
limited region of the galaxy; at least 24hr are required to obtain
full coverage of the galaxy or merger, i.e., a SNR of at least 3 in
the outer parts. We find that at least 24hr and an EE of at least 34\%
are needed to recover the merger's low surface brightness tail (see
Fig. \ref{eagle34}). We note that this EE corresponds to the EE
required to have a PCT smaller than 50\%, i.e., to have a good ``PSF
contrast'', which limits the spectroscopic coupling between adjacent
spectra (see Fig. \ref{EE_fwhm_elt}).

\begin{figure*}
\centering
\includegraphics[width=5.5cm]{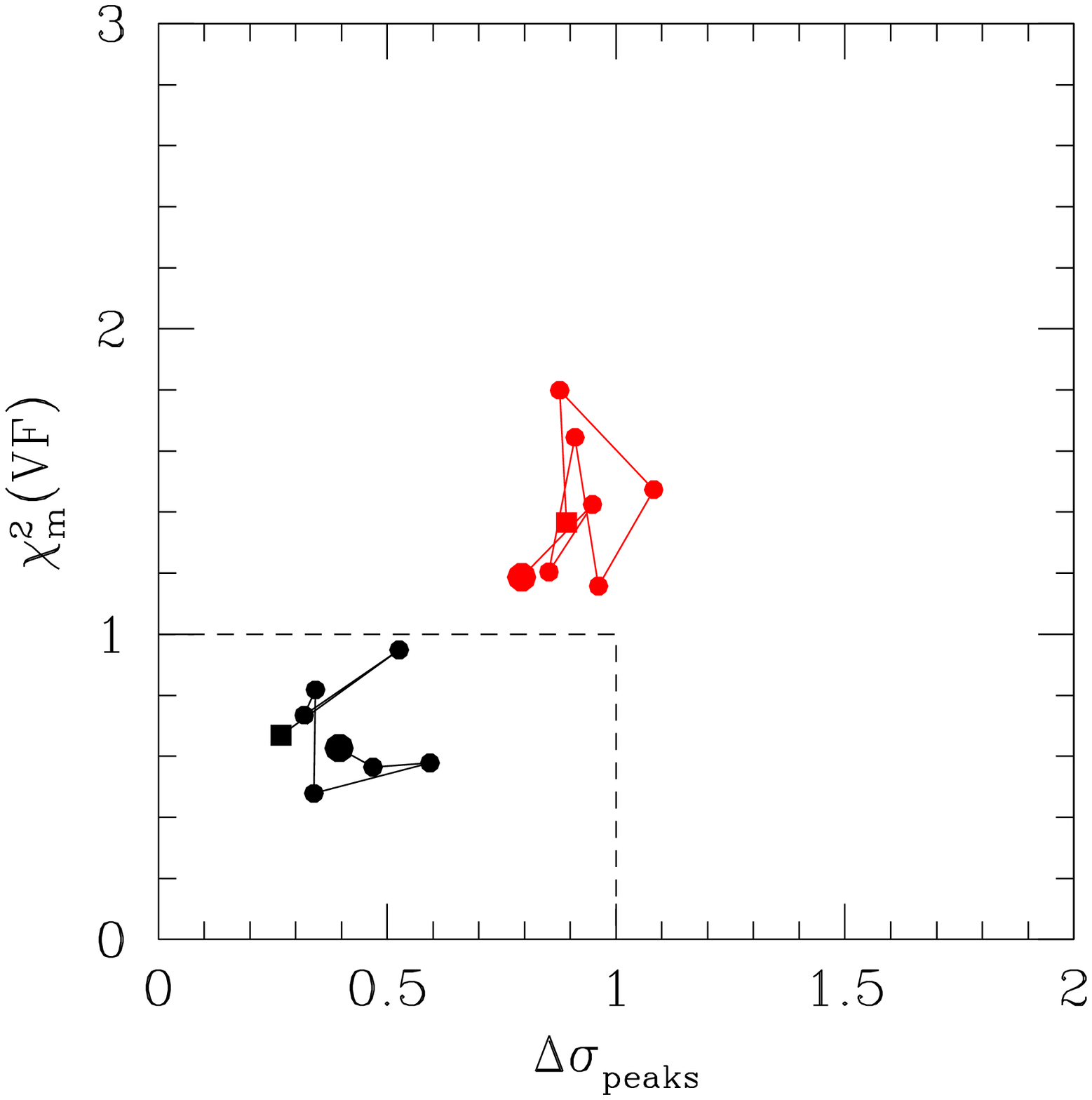}
\includegraphics[width=5.5cm]{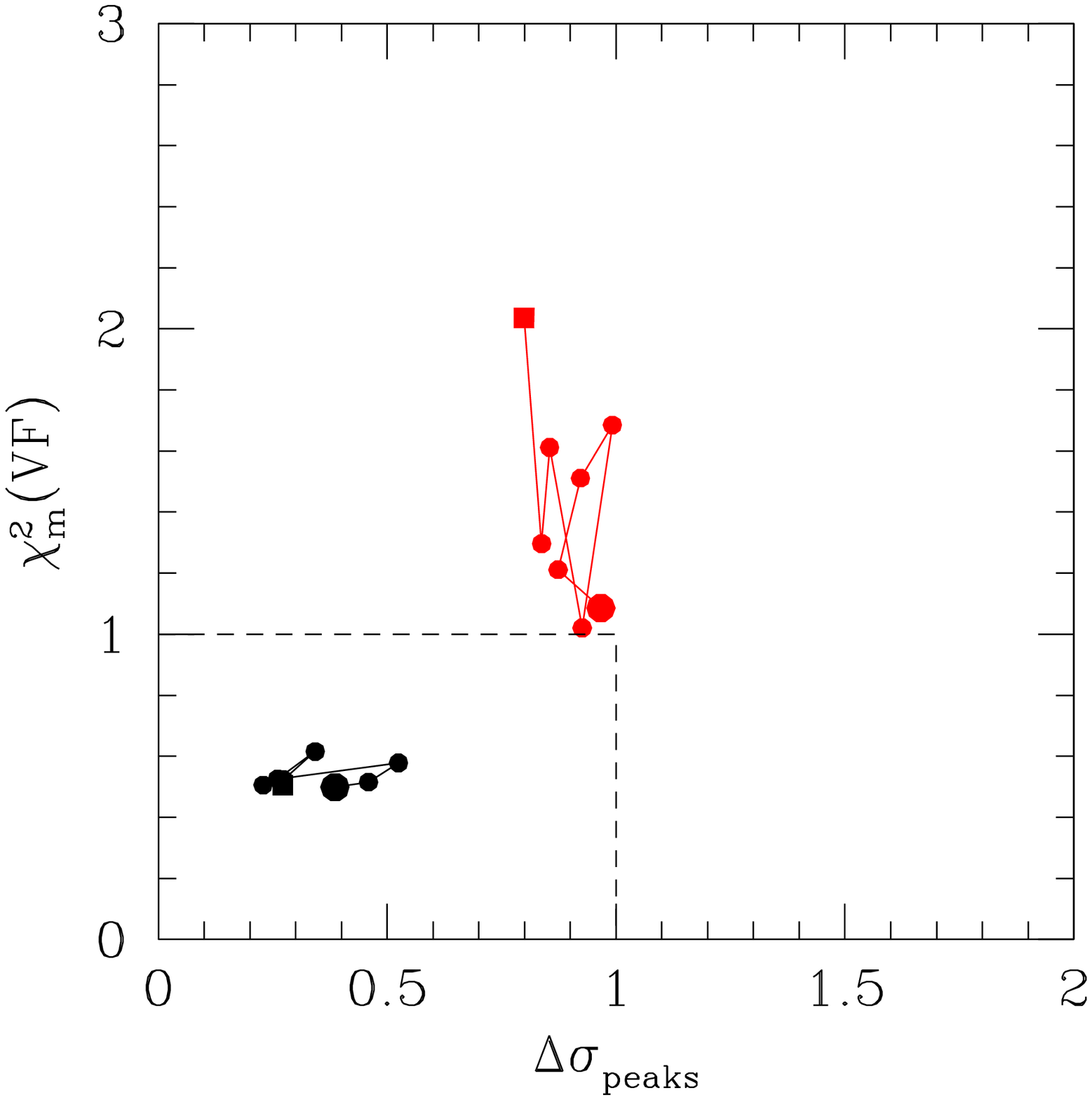}
\includegraphics[width=5.5cm]{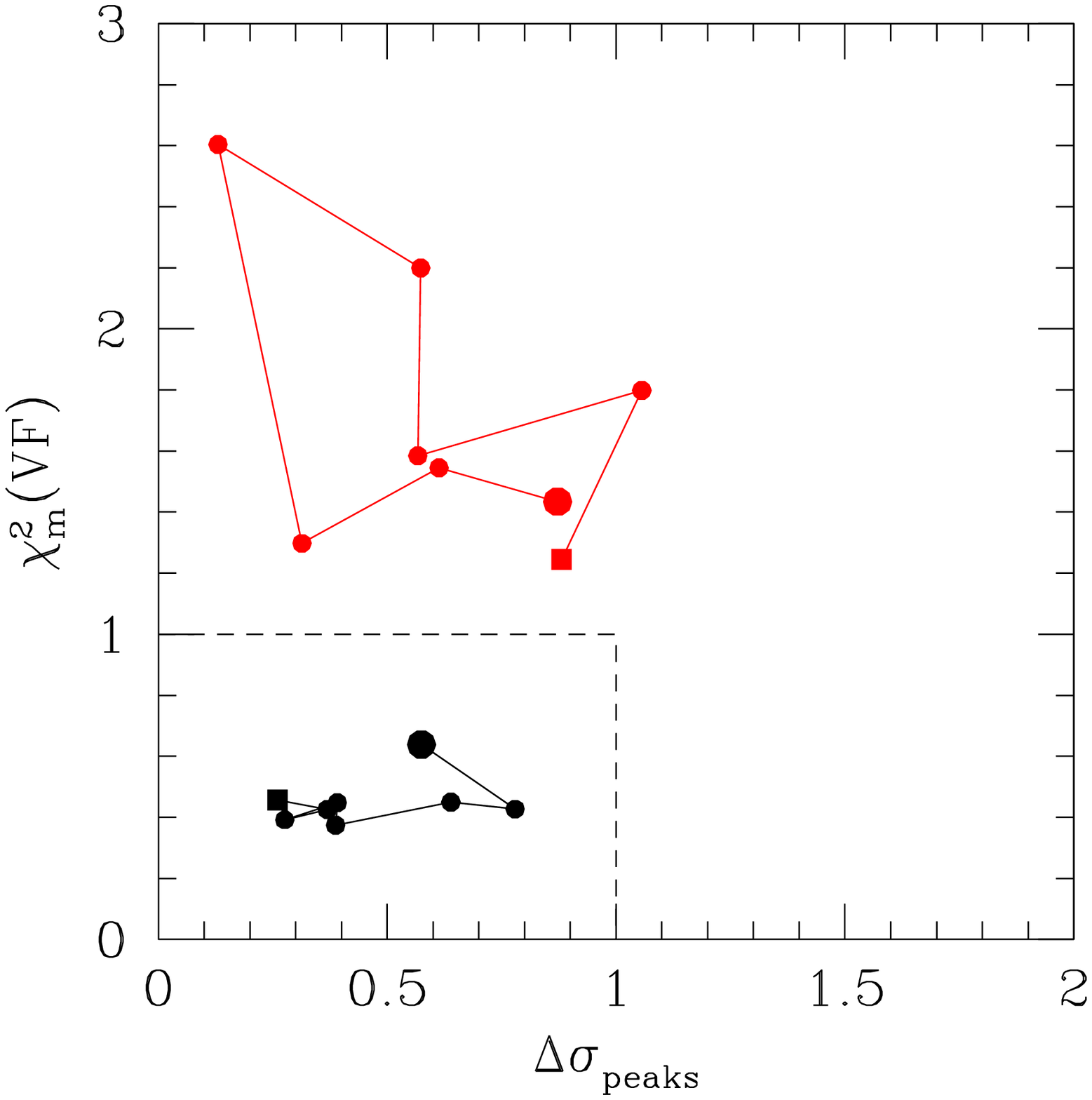}
\caption{\emph{From left to right}: Diagnostic diagram for identifying
rotating disks in EAGLE simulations of z$\sim$4 galaxies with
75mas/pixel and t$_{intg} \sim$200h, 24h, and 8h. Each point
represents a different AO correction, from the worst one (EE=22\%, see
the biggest point) to the best one (EE=47\%, see the bigger square).
The black track corresponds to the rotating disk, and the red one to
the major merger.}
\label{ELTplot}
\end{figure*}

Even when the full kinematics of the galaxy or merger is not
recovered, the total SNR is always larger that 5, down to 8hr of
exposure time. We see that this partial information is sufficient to
recover the dynamical nature of both the major merger and the rotating
disk, for a large range of EE (see Fig. \ref{ELTplot}). Although based
on a limited number of cases, this suggests that it is not mandatory
to recover the full 2D kinematics to infer the dynamical nature of
distant galaxies. The dynamical state of a galaxy is mostly reflected
in its large scale motions, rather than in its small scale
perturbations (see also \citealt{Flores06}). This is why it appears
sufficient to recover only part of the 2D information, providing that
this information is reliable enough (i.e., recovered with enough SNR).
We will discuss this point further in Sect. 8.1.

\subsection{The Design Requirements for FALCON on the VLT}

\begin{figure*}
\centering
\includegraphics[width=17cm]{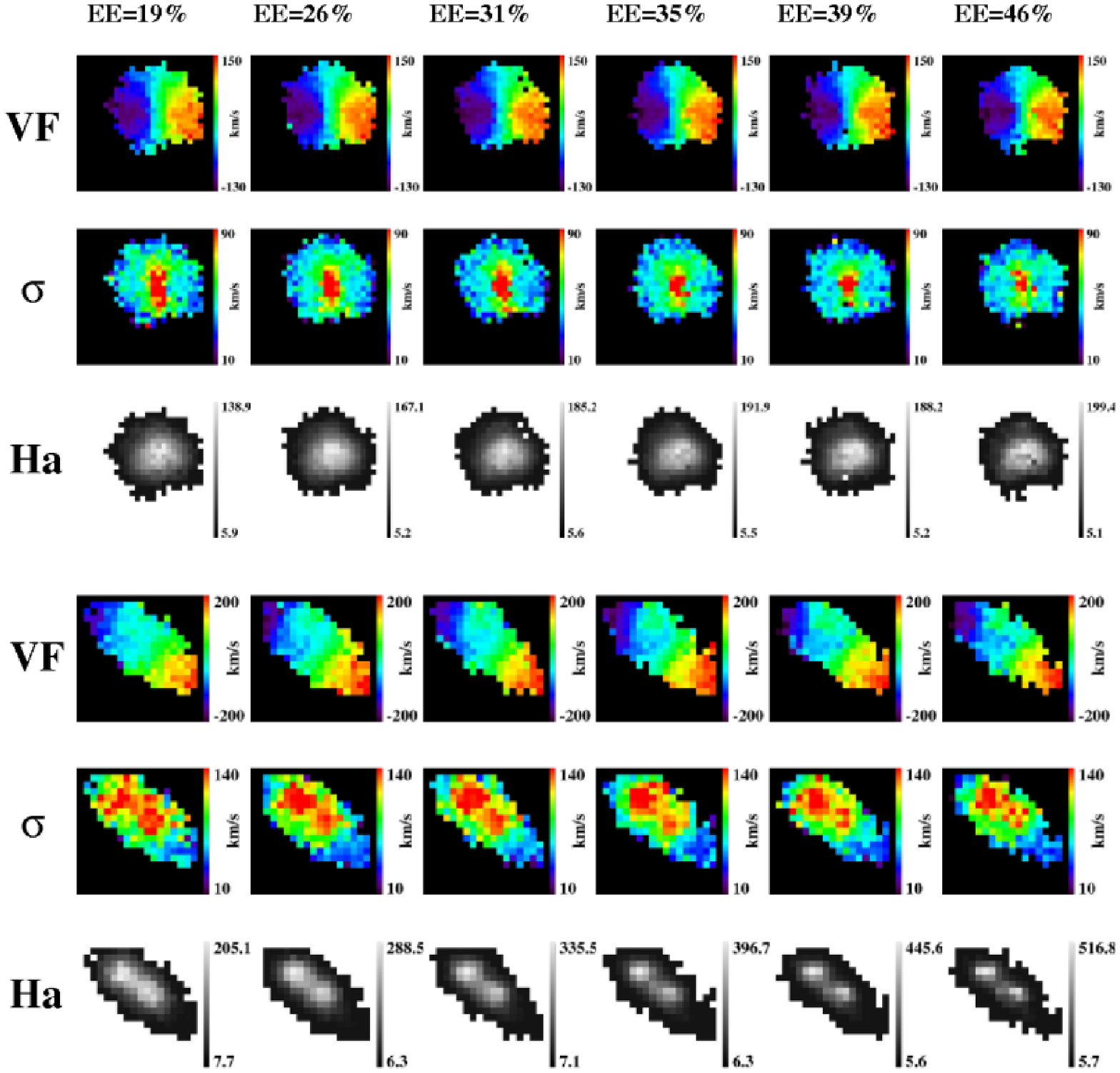}
\caption{Kinematics of a rotating disk (three top lines) and a major
merge (three lowest lines) as derived from simulations of FALCON
observations at z$\sim$1.6, with t$_{intg}$=200h. For both objects,
the velocity fields (VF), velocity dispersion maps ($\sigma$), and
emission line maps (H$\alpha$, units are in median counts per
\emph{dit}) are shown. Each column corresponds to a given EE in 0.25
arcsec, as indicated at the top.}
\label{SimFalcon200}
\end{figure*}

We proceed as described in \S 3.6, and ran a set of simulations with a
total integration time t$_{intg}$=200h (see Fig. \ref{SimFalcon200}).
To distinguish a merger from a simple rotating disk in both the
$\sigma$ and velocity maps, we find that a minimum $\sim$30\% of EE
(in 0.25 arcsec) is required. Again, this EE corresponds to a FWHM
smaller than 2 pixels (see Fig. \ref{EE_fwhm}).

\begin{figure}
\centering
\includegraphics[width=8cm]{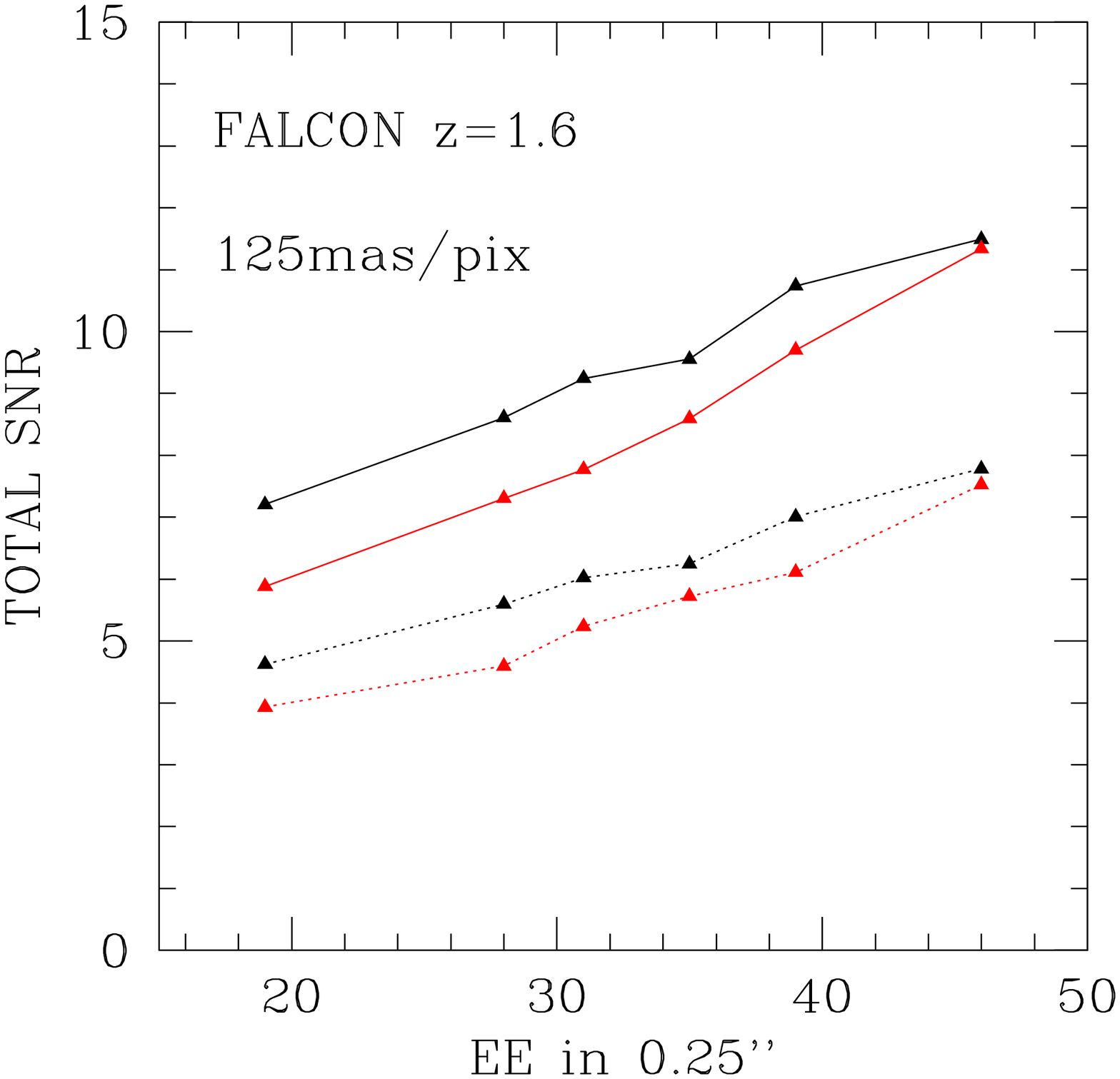}
\caption{Total flux-weighted SNR over the galaxy for FALCON
simulations with 24 and 8 hr of exposure times (respectively:
full lines, dashed lines, and dotted lines). The black curves correspond
to the rotating disk, while the red ones correspond to the major
merger.}
\label{snr_falcon}
\end{figure}

With 24hr of exposure time, the total SNR ranges between $\sim$6 and
$\sim$11, and between $\sim$4 and $\sim$8 with 8hr, depending on the EE
(see Fig. \ref{snr_falcon}). Reaching a total SNR of 5 thus requires at
least 8hr with a minimal EE of 30\%, or more than 3hr, assuming an EE
of 46\%.

\begin{figure}
\centering
\includegraphics[width=8cm]{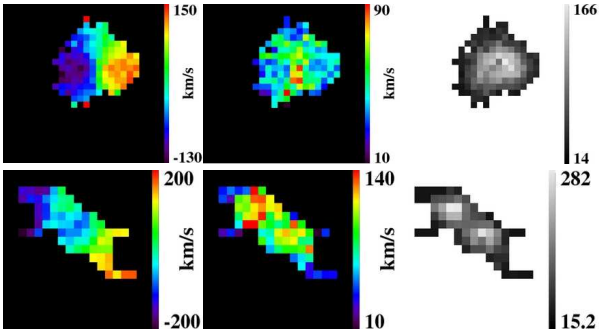}
\caption{Kinematics of a rotating disk and major merger as derived
from simulations of FALCON observations at z$\sim$1.6 (0.125
arcsec/pixel), with t$_{intg}$=24h and EE=35\%. For both objects, the
velocity fields (first column), velocity dispersion maps (second
column), and emission line maps (third column, units are in median
counts per \emph{dit}) are shown. The objects size is 2 arcsec in
diameter, which represents $\sim$0.24 kpc at z=1.6.}
\label{SimFalcon24}
\end{figure}

We find that recovering the full kinematic information requires at
least 24hr of exposure time and an EE of 35\% (see Fig.
\ref{SimFalcon24}). Again, this EE corresponds to a PCT smaller than
50\% (see Fig. \ref{EE_fwhm}), i.e., to a good ``PSF contrast''. In a
previous study, \cite{Assemat07} found that 35\% of EE was required
for detecting the H$\alpha$ emission line at a 3$\sigma$ level.
However, one can expect that for measuring higher moments of the
emission line, i.e., the first and second moments, a larger SNR will
be required. If the velocity measurement does not seem to be affected
much by the SNR, the velocity dispersion clearly suffers from a too
low SNR, since the two progenitors can hardly be directly identified
in Fig \ref{SimFalcon24}, relying on velocity dispersion maps only.

\begin{figure*}
\centering
\includegraphics[width=5.5cm]{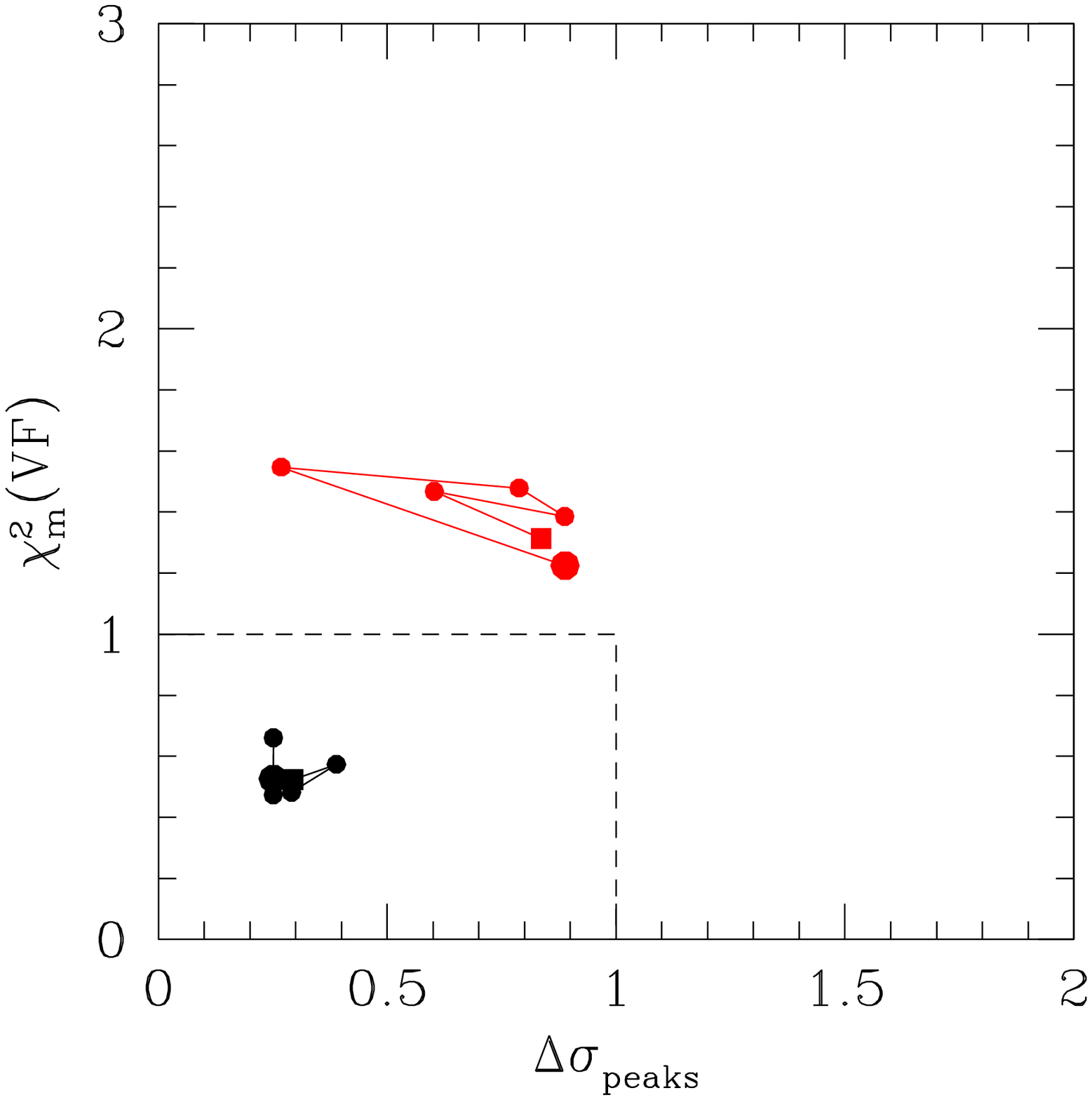}
\includegraphics[width=5.5cm]{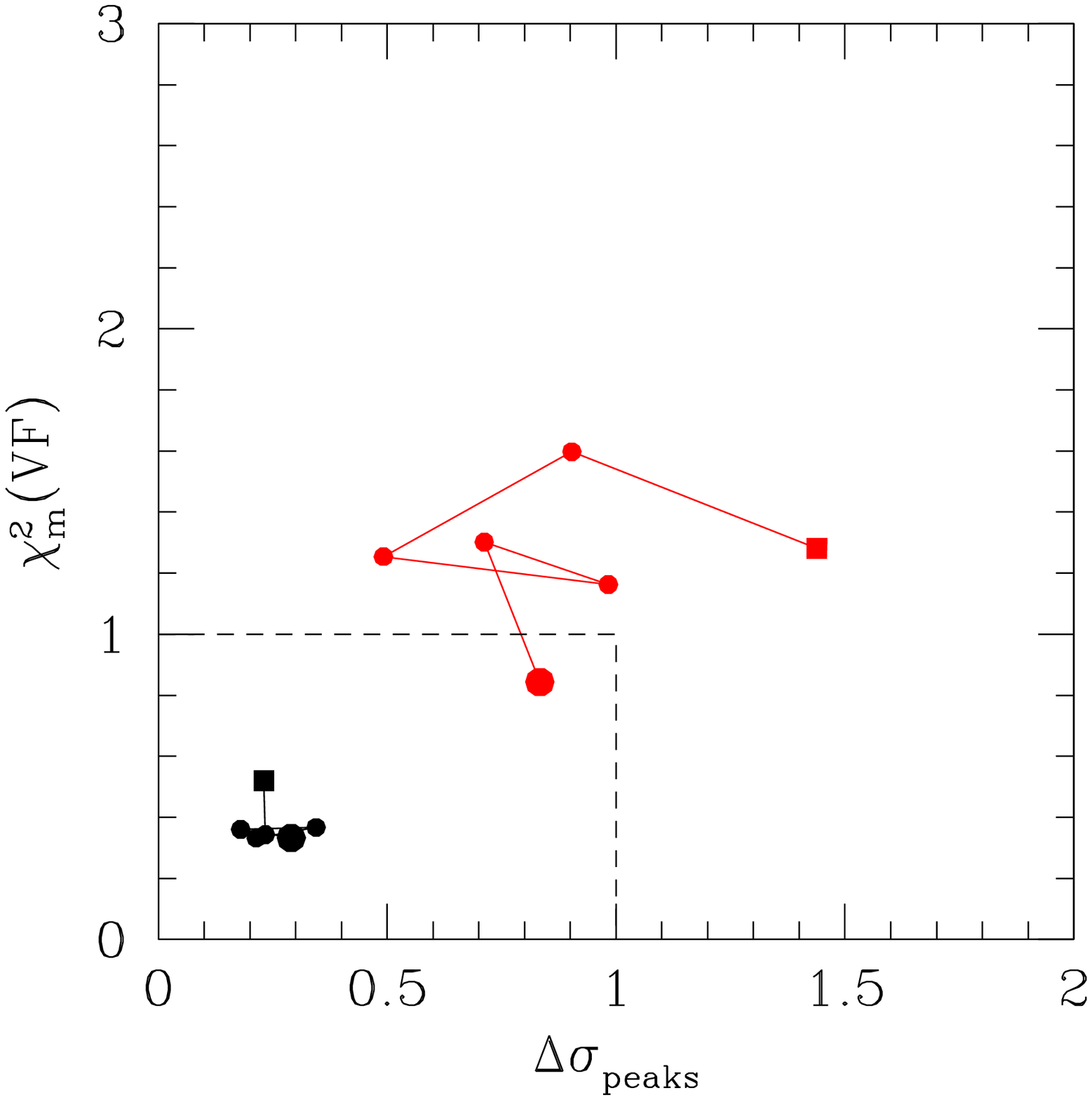}
\includegraphics[width=5.5cm]{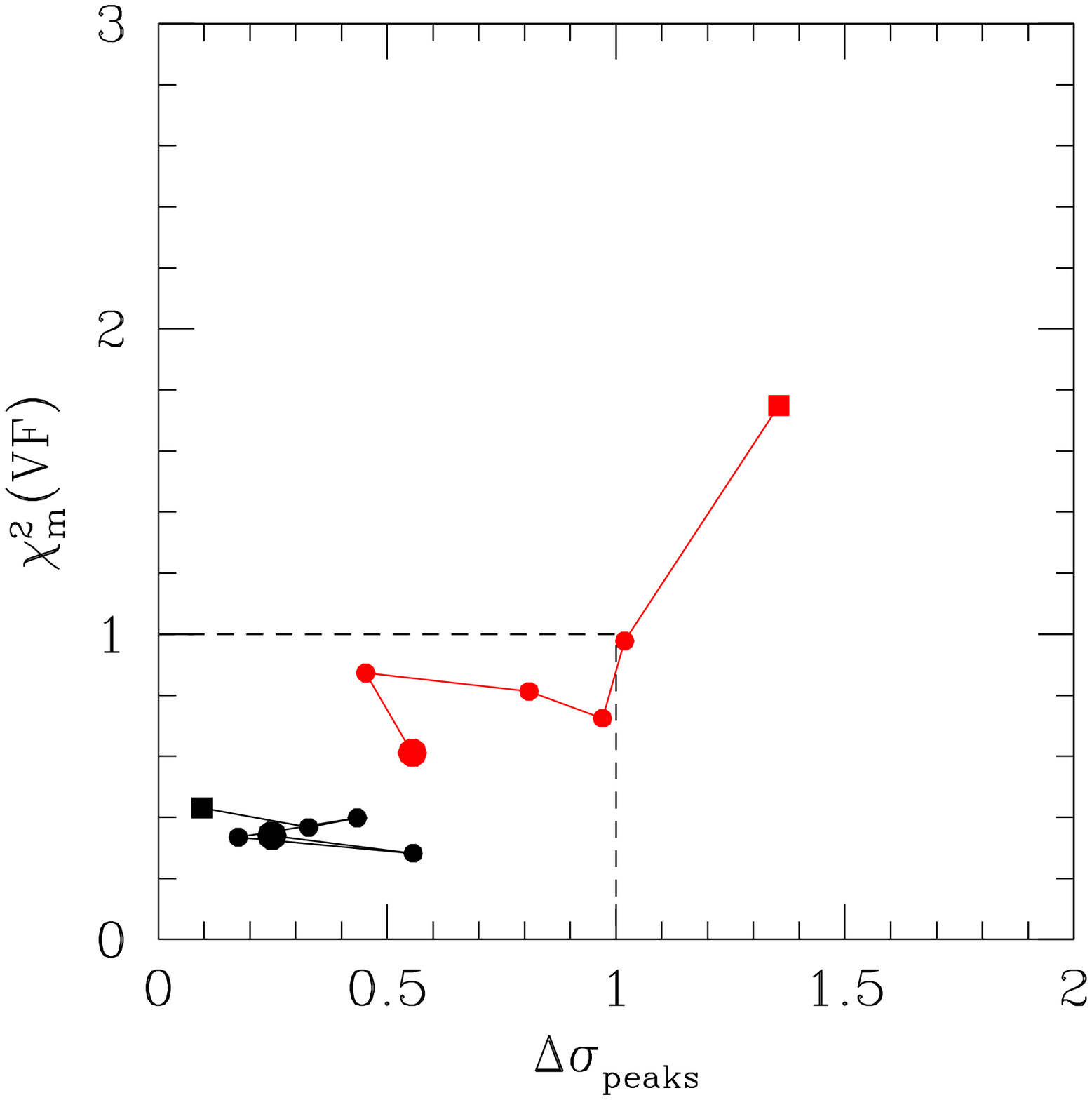}
\caption{Diagnostic diagram for identifying rotating disks in FALCON
  simulations of z$\sim$1.6 galaxies with 125mas/pixel and t$_{intg}
  \sim$200h, 24h, and 8h (\emph{from left to right}). Each point
  represents a different AO correction, from the worst one (EE=19\%,
  see the biggest point) to the best one (EE=46\%, see the bigger
  square). The black track corresponds to the rotating disk, and the
  red one to the major merger.}
\label{plotVLT}
\end{figure*}

With 8hr of integration time, it is possible to recover the true
underlying nature of dynamics of high redshift galaxies, using the
diagnostic diagram (see Fig. \ref{plotVLT}). However, a minimal EE of
26\% and 35\% are respectively necessary for integration times of 24
and 8h. However, the diagnostic diagram generally fails when the total
SNR is smaller than $\sim$5 (see Fig. \ref{snr_falcon}), confirming
once again that 5 appears to be the lowest acceptable SNR.

\section{Discussion \& Summary}

\subsection{Spatial ``scale-coupling'': Influence of the pixel size for EAGLE}

\begin{figure}
\centering
\includegraphics[width=8cm]{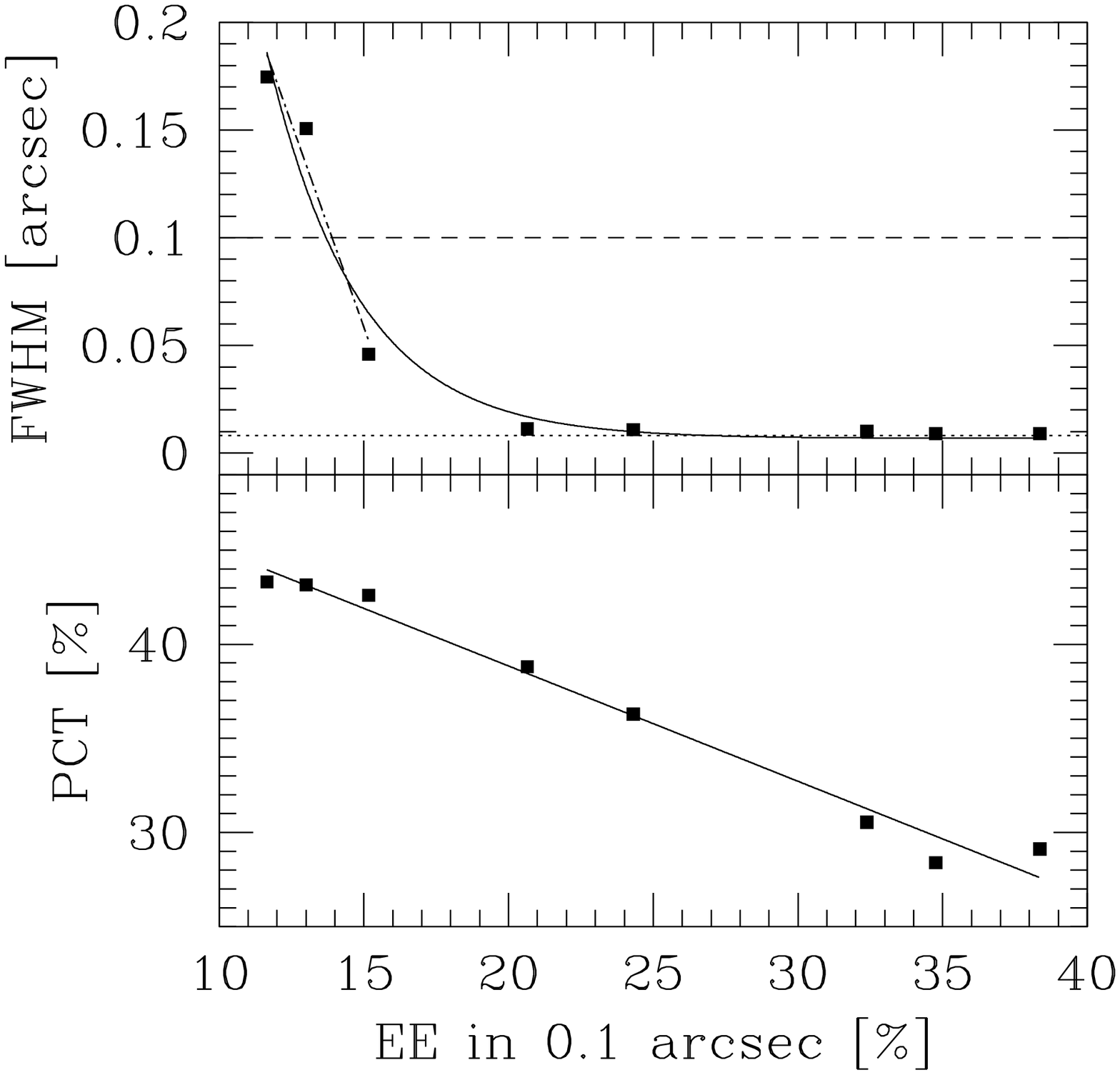}
\caption{\emph{Top:} FWHM vs. EE for the eight simulated EAGLE PSFs.
The dotted line represents the diffraction limit on a 42m ELT in the
H-band, i.e., $\lambda _{diff} (H)=0.008$ arcsec. The dashed lines shows
when the FWHM becomes lower than 0.1 arcsec, i.e., twice the pixel
size (under-sampling). \emph{Bottom:} PCT vs. EE for the eight
simulated EAGLE PSFs. The black line is a linear fit between
FWHM et EE.}
\label{EE_fwhm_elt2}
\end{figure}

\begin{figure*}
\centering
\includegraphics[width=17cm]{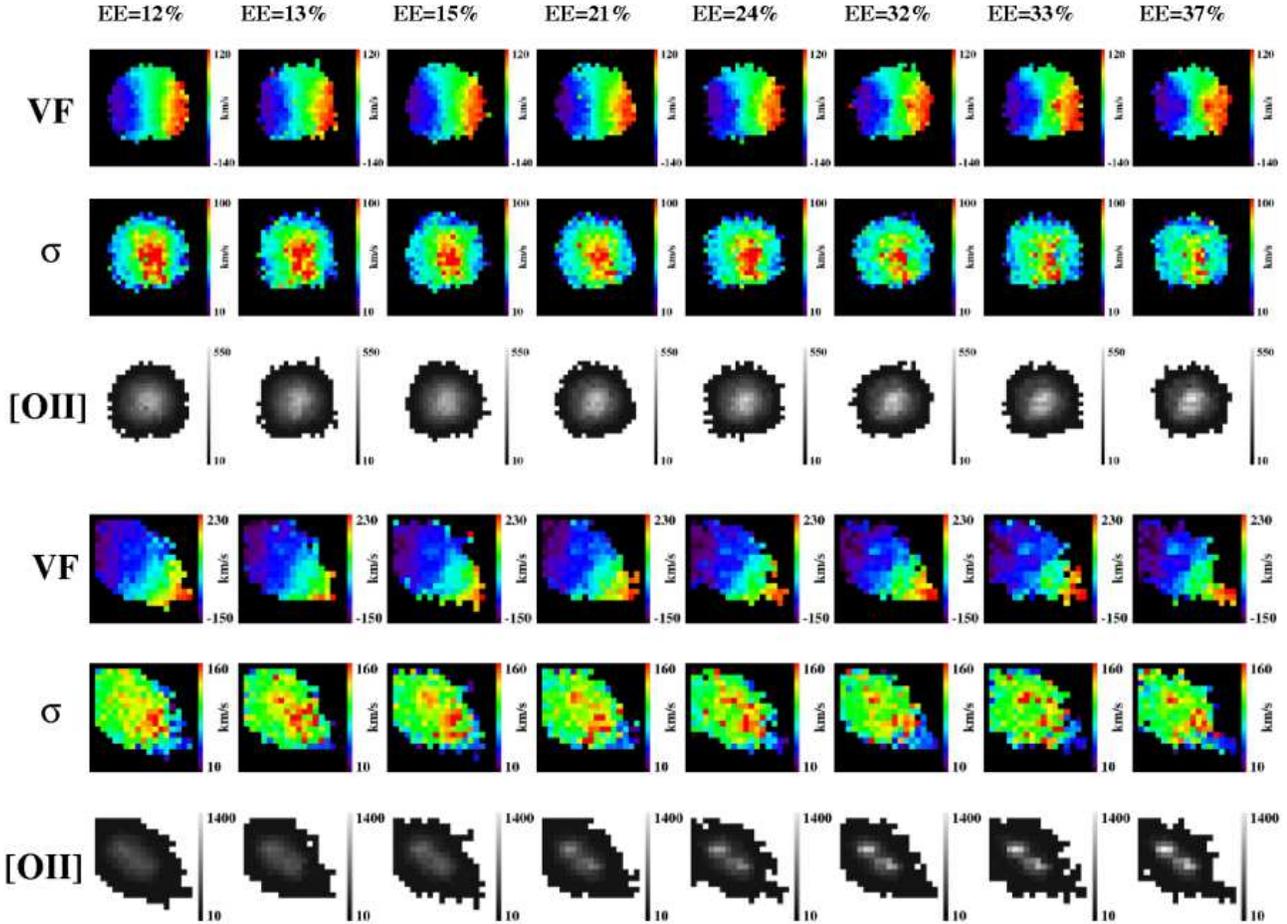}
\caption{Kinematics of a rotating disk (three top lines) and a major
merger (three lowest lines) as derived from simulations of EAGLE
observations at z$\sim$4, with 50mas/pixel and t$_{intg}$=200h. The
velocity fields (VF), the velocity dispersion maps ($\sigma$), and the
emission line flux maps ([OII], units are in median counts per
\emph{dit}) are shown. Each column corresponds to a given EE in 0.1
arcsec, as indicated at the top.}
\label{SimELT20050}
\end{figure*}

We also ran simulations of the EAGLE instrument concept for the E-ELT
with a smaller pixel scale, 50 mas (see Fig. \ref{SimELT20050}). FWHMs
and PCTs are shown in Fig. \ref{EE_fwhm_elt2}. The two progenitors of
the major merger can be clearly separated in both the $\sigma$ and the
[OII] maps (see Fig. \ref{SimELT20050}) with EE $>$15\% (in 0.1
arcsec), which again corresponds to the EE where the FWHM becomes
smaller than two pixels.

\begin{figure}
\centering
\includegraphics[width=8cm]{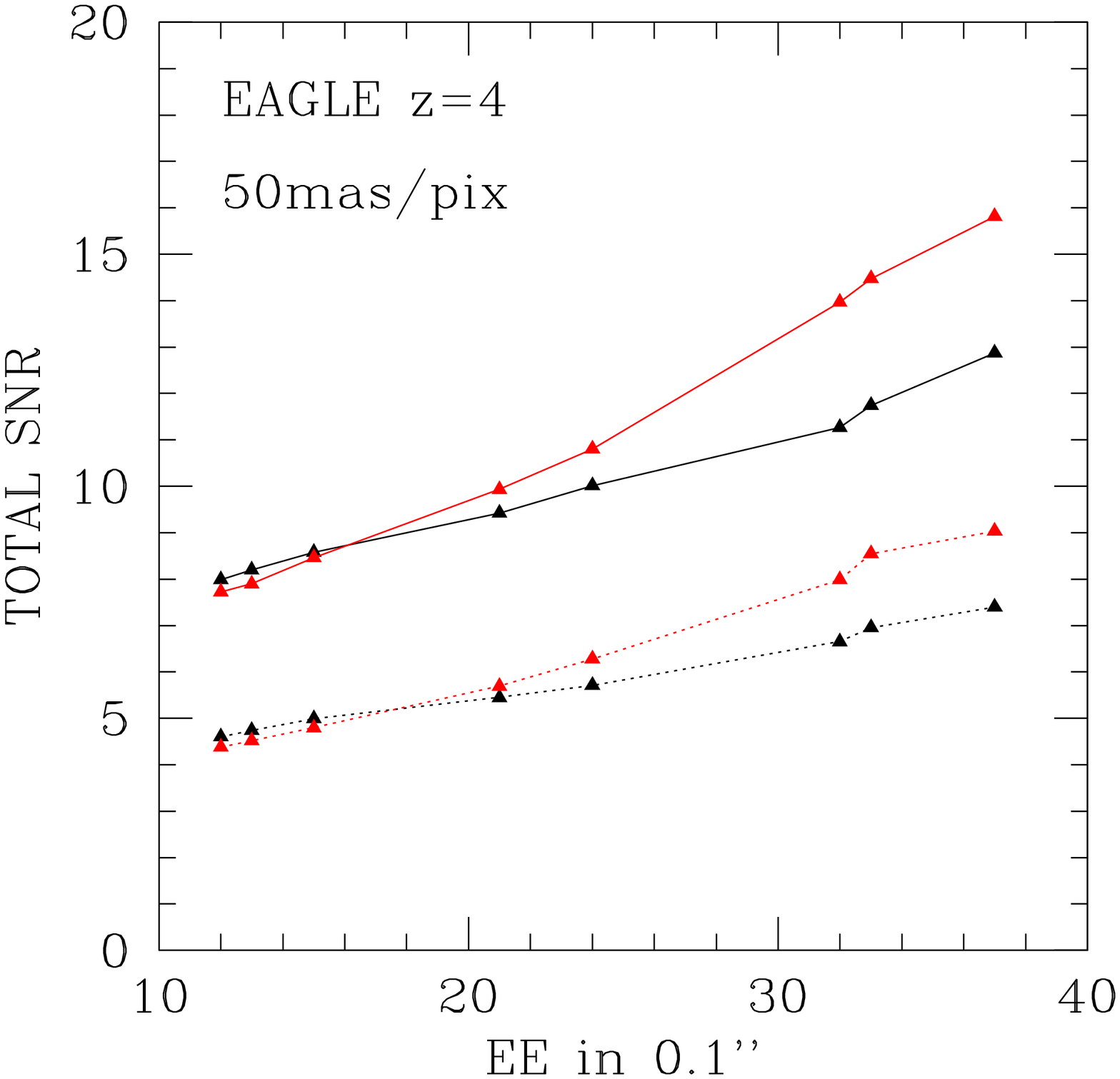}
\caption{Total flux-weighted SNR over the galaxy for EAGLE simulations
with 50mas.pix$^{-1}$ and 24 and 8 hr of exposure times (respectively:
full lines, dashed lines, and dotted lines). The black curves correspond
to the rotating disk, while the red ones correspond to the major
merger.}
\label{snr_50mas}
\end{figure}

As expected, the achieved total SNR at a given integration time and
EE, is lower with a 50 mas pixel scale than with one of 75 mas
pixel$^{-1}$ (see Fig. \ref{snr_50mas}). Reaching a minimal total SNR
of 5 requires at least 8hr of exposure time with an EE of 21\%, or at
least $\sim$3hr with a larger EE.

With 50 mas pixel$^{-1}$, we find that the low surface brightness
tidal tail of the merger can be detected in t$_{intg}$=24h with at
least 32\% of EE. In contrast to the previous cases, this EE does not
correspond to the one giving 50\% of PCT: this condition now gives a
much smaller EE (smaller than 10, see Fig. \ref{EE_fwhm_elt2}), which
does not provide enough SNR to make this distinction (Fig.
\ref{snr_50mas}). With the 75 mas pixel$^{-1}$ scale, detecting the
tidal tail required a total SNR of $\sim$15: this is roughly the total
SNR that is now provided by 32\% of EE with the 50 mas pixel$^{-1}$
scale. This is obviously related to subdividing the total flux of the
tail into smaller and smaller sampling of a structure with a
relatively smooth surface brightness distribution.

\begin{figure*}
\centering
\includegraphics[width=5.5cm]{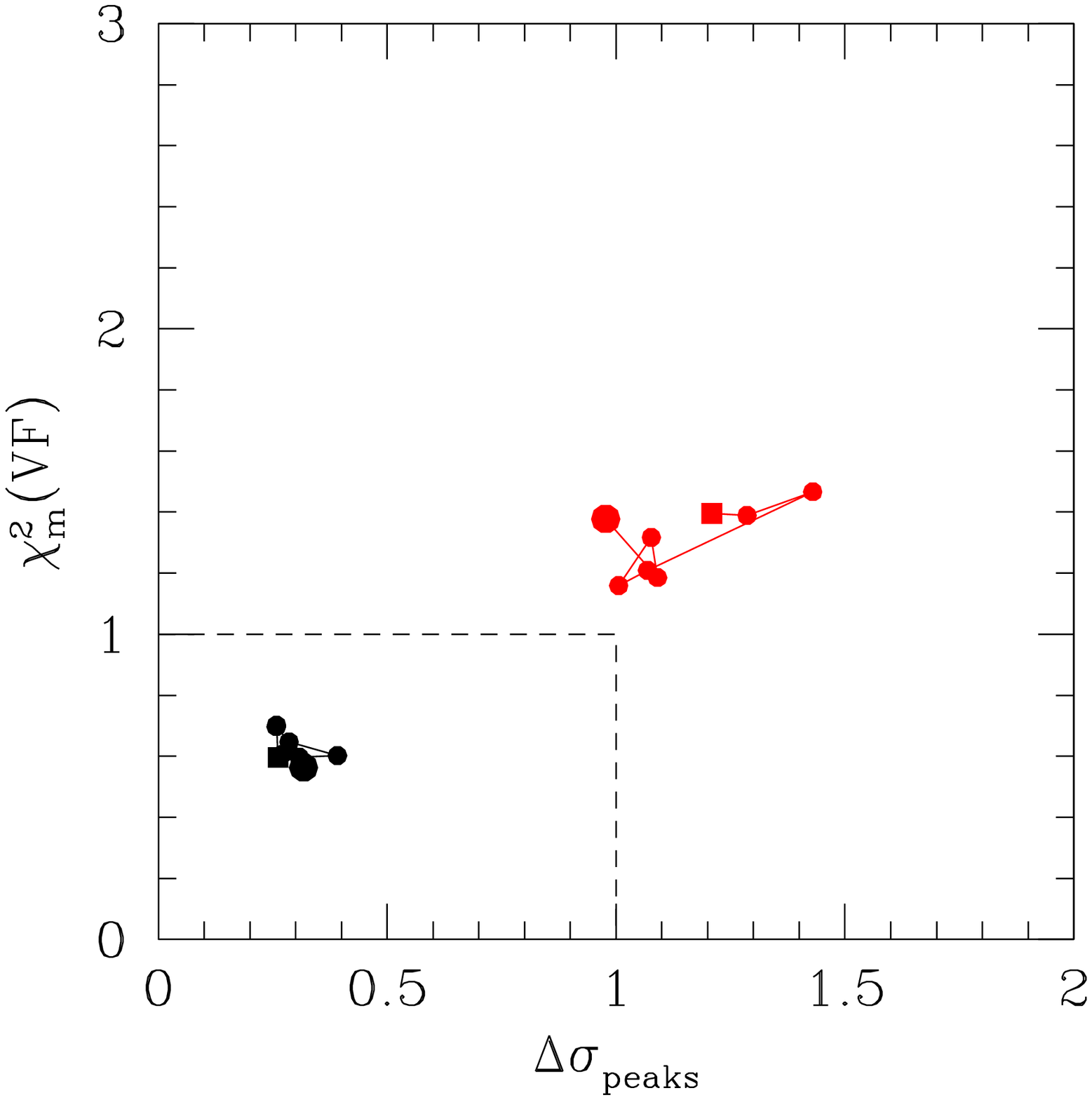}
\includegraphics[width=5.5cm]{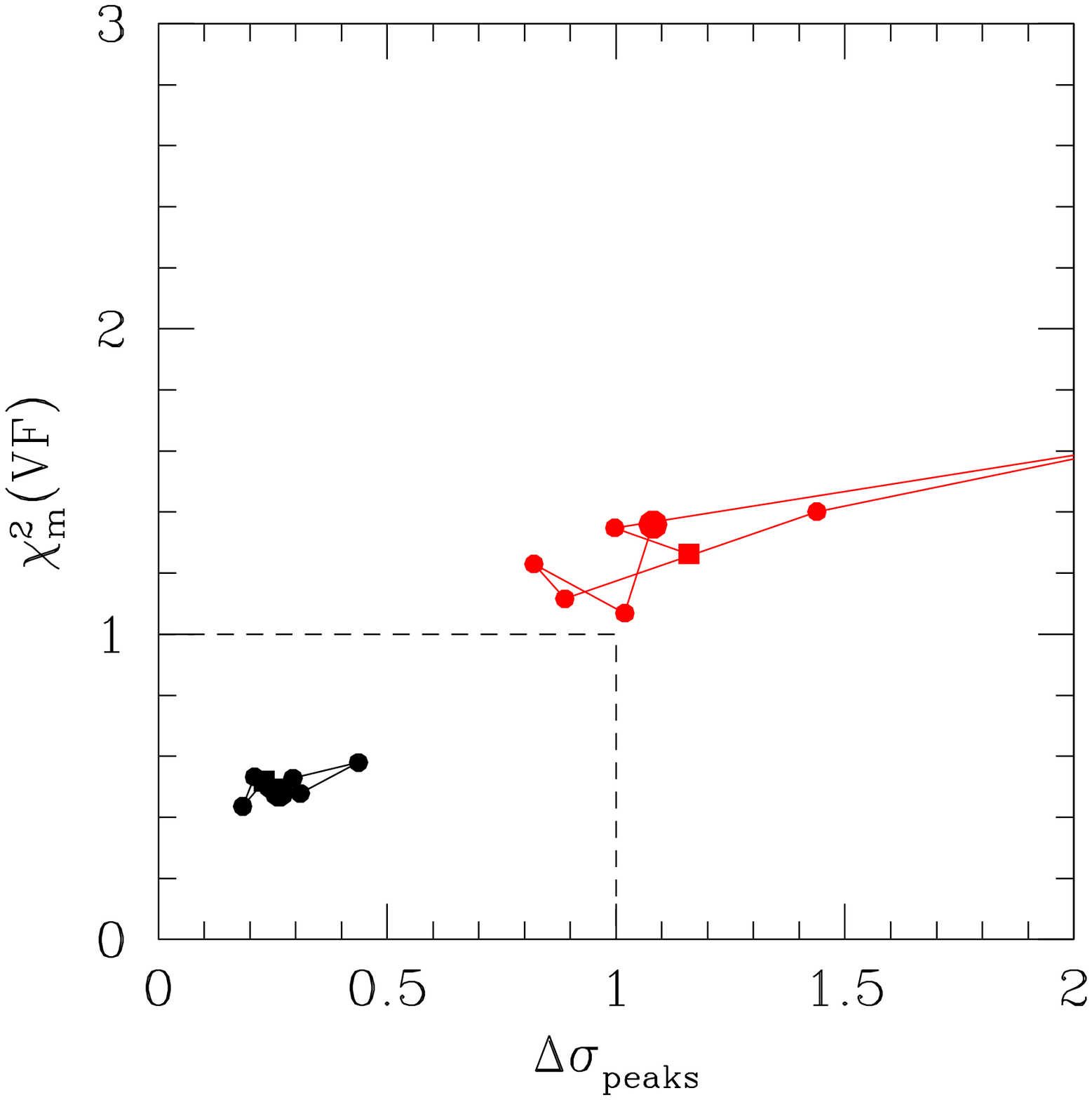}
\includegraphics[width=5.5cm]{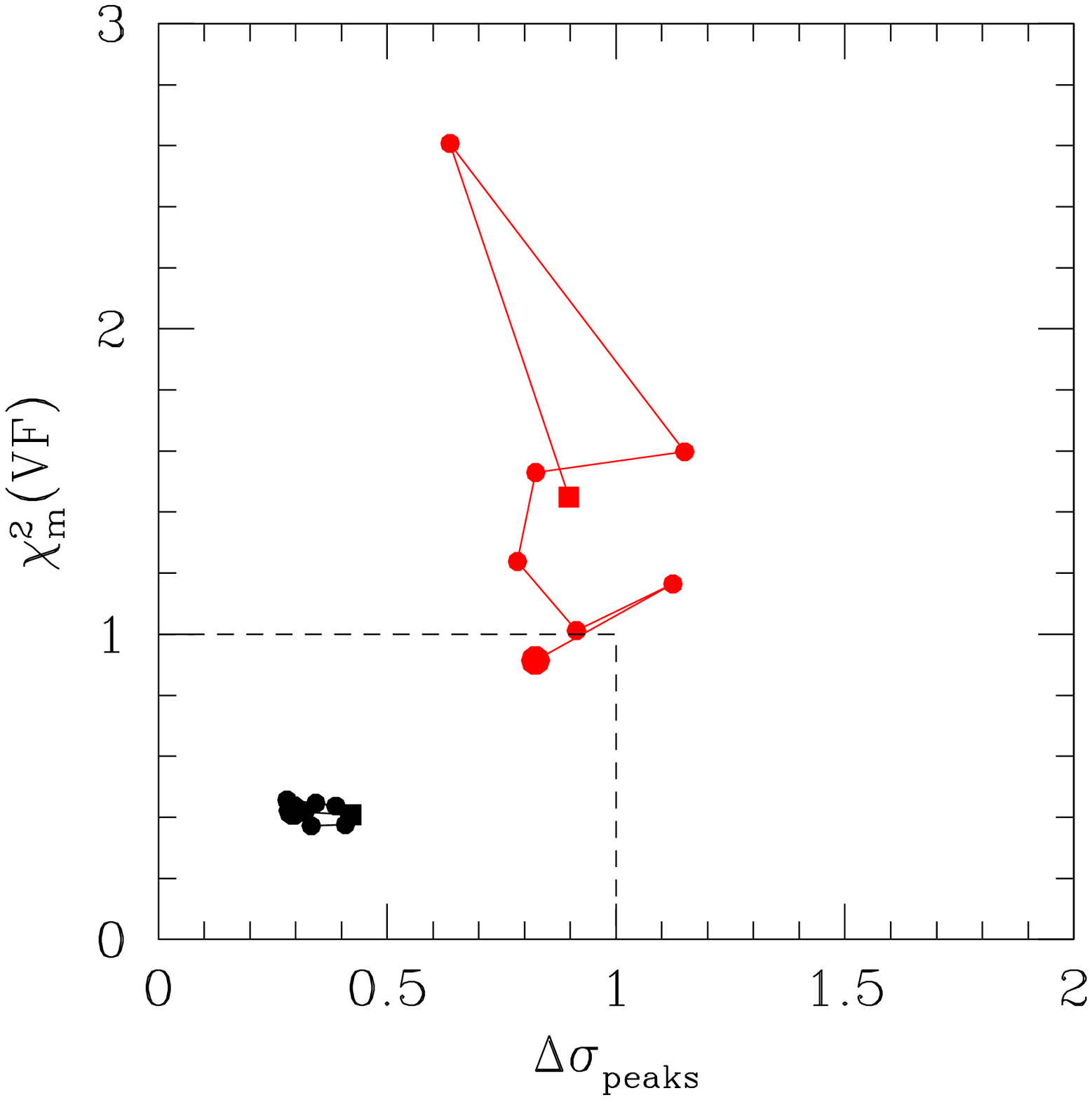}
\caption{\emph{From left to right}: Diagnostic diagram for identifying
rotating disks in EAGLE simulations of z$\sim$4 galaxies with 50 mas
pixel$^{-1}$ and t$_{intg} \sim$200h, 24h, and 8h. Each point
represents a different AO correction, from the worst one (EE=12\%, see
the biggest point) to the best one (EE=37\%, see the bigger square).
The black track corresponds to the rotating disk, and the red one to
the major merger.}
\label{ELTplot2}
\end{figure*}

Using the diagnostic diagram, it is possible to disentangle the major
merger from the rotating disk with a pixel scale of 50 mas
pixel$^{-1}$ (see Fig. \ref{ELTplot2}). Again, the diagnostic diagram
tends to fail when the total SNR is smaller than 5 (see Fig.
\ref{snr_50mas}).

The comparison of the 50 and 75 mas pixel$^{-1}$ scales reveals that
the former does not bring significant more information compared to the
latter, relatively to the goal of distinguishing between the major
merger and the rotating disk. This is related to the fact that, as
mentioned before, the dynamical state of a galaxy is reflected in its
large scale motions, rather than in its small scale perturbations.
Hence, having a very fine pixel scale does not provide any advantage
in this respect, and might limit the SNR achievable for a given
integration time. Obviously, such a conclusion depends on the spatial
scale of the kinematic feature to be recovered: if one wants to
recover smaller-scale kinematic details like, e.g., the precise
rotation curve, or to detect clumps in distant disks, a finer spatial
scale will be then preferable. This optimal ``scale-coupling'' between
the IFU pixel scale and the spatial scale of the physical feature that
one wants to recover using this IFU can be quantified by the ratio
between the size of this feature (here, the galaxy diameter, as one
wants to recover the large-scale rotation) and the size of the IFU
resolution element. At z$\sim$4, this ``scale-coupling'' is found to
be 5.3 using the 75 mas pixel$^{-1}$ scale, and 8, using the 50 mas
pixel$^{-1}$ scale. In other words, our simulations suggest that a
``scale-coupling'' of 5.3 is already large enough to recover
large-scale motions in z$\sim$4 galaxies. This is in line with 3D
observations of z$\sim$0.6 galaxies with FLAMES/GIRAFFE, which
provides a ``scale-coupling'' of about 3 \citep{Flores06,Yang07}: this
is the minimum value necessary to ensure that each side of the galaxy
is at least spatially sampled by the IFU at the Nyquist rate. Using
this minimum ratio, one finds that a pixel scale of at least $\sim$130
mas is mandatory for studying large-scale motions in z$\sim$4 galaxies
if they are simply scaled versions of nearby galaxies. On the other
hand, the smallest pixel scale that can be used for a given scientific
goal is limited by the achievable SNR at this scale, which is in turn
constrained by the ``PSF contrast'' as discussed above in Sect. 7.

\subsection{Influence of Telescope Diameter}

As discussed previously, in most cases MOAO will provide 3D
spectroscopy of distant galaxies with a spatial resolution limited by
the IFU pixel scale, because MOAO essentially provides partial AO
corrections. Everything else being equal, using a larger aperture
telescope will then only influence the integration time needed to
reach a given SNR, thanks to the larger collecting power of the
primary mirror.

\begin{figure}
\centering
\includegraphics[width=8cm]{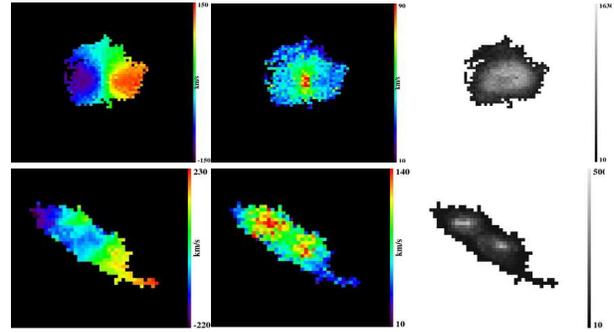}
\caption{Kinematics of a rotating disk (first line) and a major merger
(second line) as derived from EAGLE simulations at z$\sim$1.6, with
t$_{intg}$=2h and EE=24\% in 0.1 arcsec. For both objects, the
velocity fields (first column), velocity dispersion maps (second
column) and emission line maps (third column) are shown. Object size
is 2 arcsec in diameter, which represents $\sim$0.24 arcsec at z=1.6.}
\label{SimEagleFalcon}
\end{figure}

\begin{figure}
\centering
\includegraphics[width=8cm]{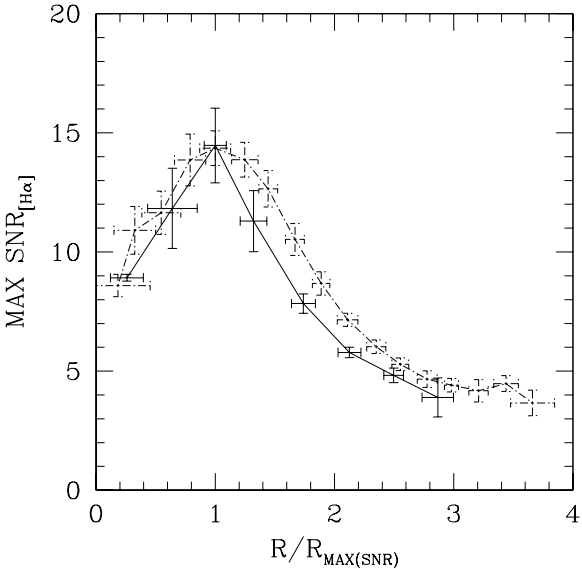}
\caption{Radius-average profile of the maximal SNR in the emission
line, obtained after 24hr of exposure time and 39\% of EE in 0.25
arcsec with FALCON (full lines), and after 8hr of exposure time and
24\% of EE in 0.1 arcsec with EAGLE (dotted lines). Error-bars account
for bootstrap uncertainties on the medians in SNR and positions within
each ring of the profile.}
\label{profsnr_comp}
\end{figure}

We show in Sect. 4, that an MOAO-fed IFS on the VLT could provide us
with improved 2D-kinematics of z$\leq$2 galaxies. In Fig.
\ref{SimEagleFalcon}, we show what the maps of the same galaxy would
look like if observed with EAGLE on the E-ELT, using 50 mas
pixel$^{-1}$ and t$_{intg}$=2h. Observations, with 24\% of EE in 0.1
arcsec, provide a radial SNR profile comparable to the one obtained
with FALCON after 24hr of exposure time and EE=39\% (see Fig.
\ref{profsnr_comp}). The E-ELT, equipped with an multi-object IFS,
even with a low EE, would represent a huge improvement over 8-meter
class telescopes, even equipped with future-generation MOAO-fed
instruments. In return, the VLT, equipped with an MOAO-fed
spectrograph would provide us with an important advancement in our
knowledge of distant galaxies, increasing the number of spatial
element of resolution from typically 2-3 to 13-20. This will be
possible at the VLT with the Laser Guide Star Facility and SINFONI
(pixel scale of 50 mas). This is already available with OSIRIS at Keck
(with 20 to 100 mas pix$^{-1}$ AO-corrected; see
\citealt{Law07,Wright07}). Other VLT second-generation NIR
spectrographs, such as KMOS, will on the other hand allow us to gather
further information on kinematics of 24 distant objects
simultaneously, which is a great advantage in obtaining the
statistical properties. However, it only has seeing limited
performance over a 0.2 arcsec pixel$^{-1}$ scale. High spatial
resolution integral field spectroscopy at the VLT will remain limited
to single object observations. The number density of objects at
z$\sim$1.4-2.5 is expected to be $\sim$ 9 per square arcmin, down to
$I_{AB} \sim$25 \citep{Steidel04}. KMOS will undoubtedly identify
interesting objects amount the ensemble of z$\approx$2 galaxies with
which to observe in more detail with AO-fed integral field
spectrographs on 8-m class telescopes. However, an MOAO-fed
spectrograph on currently available or a future extremely large
telescope would optimally take advantage of these high target
densities.

\subsection{Summary: High Level Design Requirements for MOAO systems} 

We first presented simulations of z$\sim$4 galaxies with very large
SNRs, to focus on spatial smoothing effects due to partial corrections
of the atmospheric turbulence by the MOAO system. In all cases, we
find that the EE required to clearly identify the two components of
the major merger considered in our simulations roughly corresponds the
EE that provides us with a FWHM smaller than 2 pixels. For z=4
galaxies and using a 75mas pixel scale, this corresponds to an EE of
$\sim$25\% in 150 mas, while this condition translates into 15\% in
100 mas with a 50mas pixel$^{-1}$ scale. At z=1.6, an EE of at least
$\sim$30\% within 0.25 arcsec is required, using a 0.125 arcsec
pixel$^{-1}$ scale. Considering the spatial resolution only, (i.e.,
with an infinite SNR), these specifications are indications of the
minimum we will realistically require to recover spatially resolved
kinematic information on distant galaxies.

Considering the impact of signal to noise, if one wants to recover the
full 2D kinematics of z=4 galaxies, at least 24 hr of integration
time, and 34\% of EE (in 0.15 arcsec, using the 75 mas pixel$^{-1}$
scale) are needed (or 32\% in 0.1 arcsec using the 50 mas pixel$^{-1}$
scale). For z=1.6 FALCON simulations, at least 24 hr and 35\% of EE in
0.25 arcsec are needed for recovering the full 2D kinematics in our
simulations. When the SNR provided by the MOAO system is large enough,
these EE roughly corresponds to a PCT smaller than 50\%. This provides
a good ``PSF constrast'' at a given spatial scale, which avoids
significant amount of polluting light from adjacent spectra.

At lower SNR, the required EE naturally depends on the surface
brightness of the faintest regions to be detected, as well as on the
adopted pixel scale. The choice of the latter strongly depends on the
scientific goal and the size of the features to be recovered by 3D
spectroscopy: a good ``scale-coupling'' between them must provide a
least a few spatial element of resolutions per important scale.
Because the dynamical state of galaxies is mostly reflected in their
large-scale motions, this means that only a few spatial element of
resolution are needed per galaxy diameter: we performed additional
simulations (not shown here) using another rotating disk with a less
steep rotation curve (UGC6778, see \citealt{Garrido02}), and found no
quantitative difference in our results using this other rotating disk.
Although based on a limited number of simulations, we find that indeed
a spatial resolution element of 150 mas (pixel size of 75 mas) seems
to be a good choice for recovering large scale motions, providing
relatively large SNR with a ``scale-coupling'' $\sim$5.3 element of
resolution per galaxy diameter (0.8 arcsec). At the 50mas pixel$^{-1}$
scale, a 42 m ELT will typically need 8 hr of exposure time to reach a
SNR of at least 5 over a galaxy. Such a SNR allows us to use the
diagnostic diagram as a useful tool to recover the dynamical nature of
rotating disks and major mergers. Because of the much smaller
collecting area of the VLT, this exercise is limited to 8 hr with at
least 35\% of EE in 0.25 arcsec, for z=1.6 galaxies. In this case, the
``coupling factor'' is 4. These results are consistent with previous
studies at lower redshift (z$\sim$0.6), where the dynamical nature of
galaxies have been recovered using GIRAFFE, with only 3 or so
resolution elements (at 0.52 arcsec pixel$^{-1}$) across each galaxy.

\section{Conclusion}
We have developed software capable of end-to-end simulations of
integral field spectrograph fed by an MOAO system. We have used this
software to investigate a limited number of simulations in H-band, in
order to give insights into how such MOAO-fed systems should perform
on the VLT and the E-ELT. By requiring that the instrument is able to
distinguish between a rotating disk and a major merger, we were able
to constrain the required Ensquared Energy to be able to recover their
large-scale motions. Separating the progenitors of the major merger
requires only modest EE (15-26\% at z=4 in a sample of 100-150mas;
30\% at z=1.6 in 0.25 arcsec), but long exposure times ($\sim$24hr).
Higher EEs are generally needed to recover the full 2D kinematics in
our simulations, typically 35\%. Provided that the total SNR is large
enough over the galaxy (typically $\sim$ 5), it is possible to use
more sophisticated methods, such as the diagnostic diagram introduced
here, to distinguish between both types of objects, even at relatively
low integration time for such distant objects (8 hrs). To generalize
these results beyond what we described here, we have related these
specifications to the concept of ``PSF contrast'', which is the amount
of polluting light from adjacent spectra and which drives the minimal
EE at a given spatial scale. The choice of the latter is driven mostly
by the ``scale-coupling'' which is the relative size of the IFU pixel
and the size of the features that the observe wishes to recovered in
the data cube. Distinguishing between a major merger and a disk, for
example, can be largely done by investigating the large scale motions
within a system, and a relatively coarse spatial resolution appears
then to be sufficient. Given this situation, a pixel scale of 50-75
mas seems to be a relatively good choice, since it provides sufficient
SNR in less time than a finer sampling and also has the advantage of
relaxing the MOAO system requirements. More stringent requirements can
be set by attempting to resolve and investigate structure in high
redshift galaxies and that will be simulated in subsequent papers.

\section*{acknowledgments}
We are especially indebted to T.J. Cox who has provided us with the
hydro-dynamical simulations of merging galaxies, as well as with P.
Amram et B. Epinat who have provided us with kinematical data of local
galaxies from the GHASP survey. We wish to thank an anonymous referee
for very useful comments and suggestions. M.P. wishes to thank R.
Gilmozzi for financial support at ESO-Garching, where this work has
been finalized. This work has benefited from interesting discussions
with the WFSPEC-EAGLE team, and more especially with E. Gendron, P.
Laporte, and F. Hammer, as well as with the E-ELT Science Working
Group. We thank H. Schnetler for a very careful reading of the paper.
This work also received the support of PHASE, the high angular
resolution partnership between ONERA, Observatoire de Paris, CNRS and
University Denis Diderot Paris 7.

\label{lastpage}


\begin{thebibliography}{}

\bibitem[Amram et al.(2002)]{Amram02} Amram, P., Adami, C., Balkowski, C., et
  al. 2002, Ap\&SS, 281, 393
\bibitem[Ass\'emat et al.(2007)]{Assemat07} Ass\'emat, F., Gendron, E., \&
  Hammer, F. 2007, MNRAS, 376, 287
\bibitem[Beauvais \& Bothun(1999)]{Beauvais99} Beauvais, C., \&
  Bothun, G. 1999, ApJS, 125, 99
\bibitem[Bertin \& Arnouts(1996)]{Bertin96} Bertin, E., \& Arnouts, S. 1996,
  A\&AS, 117, 393
\bibitem[Beuzit et al.(2006)]{Beuzit06} Beuzit, J.L., Feldt, M.,
  Dohlen, K., et al. 2006, Msngr, 125, 29
\bibitem[Bosma(1978)]{Bosma78} Bosma, A. 1978, PhD Thesis dissertation
\bibitem[Bouwens et al.(2004)]{Bouwens04} Bouwens, R.J., Illingworth, G.D.,
  Blakeslee, J.P., et al. 2004, ApJ, 611, 1
\bibitem[Cox et al.(2004)]{Cox04} Cox, T.J., Primack, J., Jonsson, P., et al.
  2004, ApJ, 607, 87
\bibitem[Cox et al.(2006)]{Cox06} Cox, T.J., Jonsson, P., Primack, J., et al.
  2006, MNRAS, 373, 1013
\bibitem[Dahlen et al.(2007)]{Dahlen07} Dahlen, T., Mobasher, B.,
  Dickinson, M., et al. 2007, ApJ, 654, 172
\bibitem[Erb et al.(2006)]{Erb06} Erb, D.K., Steidel, C.C., Shapley, et al.
2006, ApJ, 647, 128
\bibitem[Ferguson et al.(2004)]{Ferguson04} Ferguson, H.C., Dickinson,
  M., Giavalisco, M. et al. 2004, ApJ, 600, 107
\bibitem[Finger et al.(2006)]{Finger06} Finger, G., Garnett, J., Bezawada, N.,
  et al. 2006, Nuclear Instruments \& Methods in Physics Research A,
  565, 241
\bibitem[Flores et al.(2006)]{Flores06} Flores, H., Hammer, F., Puech, M., et
  al. 2006, A\&A, 455, 107
\bibitem[F\"orster-Schreiber et al.(2006)]{Forster06} F\"orster-Schreiber, N.,
  Genzel, R., Lehnert, M.D., et al. 2006, ApJ, 645, 1062
\bibitem[Fried(1981)]{Fried81} Fried, D.L. 1981, JOSAA, 72, 52
\bibitem[Fuentes-Carrera et al.(2004)]{Fuentes04} Fuentes-Carrera, I.,
  Rosado, M. Amram, P., et al. 2004, A\&A, 415, 451
\bibitem[Fusco et al.(1999)]{Fusco99} Fusco, T., Conan, J.-M., Michau, V., et
  al. 1999, SPIE Proc. Vol. 3763, 125
\bibitem[Fusco et al.(2006)]{Fusco06} Fusco, T., Rousset, G., Sauvage,
  J.-F., et al., 2006, Opt. Expr., 14, 7515
\bibitem[Garrido et al.(2002)]{Garrido02} Garrido, O., Marcelin, M.,
  Amram, P. et al. 2002, A\&A, 387, 821
\bibitem[Goncharov et al.(2007)]{Goncharov07} Goncharov, A.V.,
Devaney, N., \& Dainty, C. 2007, Opt. Expr. 15, 1534
\bibitem[Hammer et al.(1997)]{Hammer97} Hammer, F., Flores, H., Lilly, S.J., et
  al. 1997, ApJ, 481, 49
\bibitem[Hammer et al.(2002)]{Hammer02} Hammer, F., Say\`ede, F., Gendron, E.,
  et al. 2002, Proc. of the ESO workshop Scientific Drivers for ESO
  Future VLT/VLTI Instrumentation, 139
\bibitem[Hammer et al.(2004)]{Hammer04} Hammer, F., Puech, M.,
  Ass\'emat, F., et al. 2004, Proc. SPIE Vol. 5382, 727
\bibitem[Hammer et al.(2005)]{Hammer05} Hammer, F., Flores, H., Elbaz, D., et
  al. 2005, A\&A, 430, 115
\bibitem[Law et al.(2007)]{Law07} Law, D.R., Steidel, C.C., Erb, D.K., et al.
  2007, ApJ, accepted, astro-ph/0707.3634
\bibitem[Mannucci et al.(2001)]{Mannucci01} Mannucci, F., Basile, F.,
  Poggianti, B.M., et al. 2001, MNRAS, 326, 745
\bibitem[Moretto et al.(2006)]{Moretto06} Moretto, G., Bacon, R.,
  Cuby, J.-G., et al. 2006, Proc. SPIE Vol. 6269, 76
\bibitem[Neichel et al.(2006)]{Neichel06} Neichel, B., Conan, J.-M., Fusco, T.,
  et al. 2006, Proc. SPIE Vol. 6272, 58
\bibitem[Puech \& Say\`ede(2004)]{Puech04} Puech, M., \& Say\`ede, F.,
Proc. SPIE Vol. 5492, 303
\bibitem[Puech et al.(2005)]{Puech05} Puech, M., Chemla, F., Laporte, P., et
  al. 2005, Proc. SPIE Vol. 5903, 272
\bibitem[Puech et al.(2006a)]{Puech06a} Puech, M., Hammer, F., Flores, H., et
  al. 2006, A\&A, 455, 119
\bibitem[Puech et al.(2006b)]{Puech06b} Puech, M., Flores, H., Hammer, F., et
  al. 2006, A\&A, 455, 131
\bibitem[Puech et al.(2008)]{Puech08}Puech, M., Flores, H., Hammer, F., et
  al. 2008, A\&A accepted, astro-ph/0803.3002
\bibitem[Reddy et al.(2006)]{Reddy06} Reddy, N.A., Steidel, C.C., Erb, D.K., et
  al. 2006, ApJ, 653, 1004
%\bibitem[Rigaut et al.(1992)]{Rigaut92} Rigaut, F. \& Gendron, E. 1992, A\&A,
%  261, 677
\bibitem[Roddier(1981)]{Roddier81} Roddier, F. 1981, Progress in
Optics, 19, 281
\bibitem[Sarzi et al.(2006)]{Sarzi06} Sarzi, M., Falcon-Barroso, J.,
  Davies, R.L., et al. 2006, MNRAS, 366, 1151
\bibitem[Semelin \& Combes(2005)]{Semelin05} Semelin, B., \& Combes, F. 2005,
  A\&A, 441, 55
\bibitem[Steidel et al.(2004)]{Steidel04} Steidel, C.C., Shapley,
  A.E., Pettini, M., et al. 2004, ApJ, 604, 534
\bibitem[Tallon \& Foy(1990)]{Tallon90} Tallon, M. \& Foy, R. 1990, A\&A, 235, 549
%\bibitem[Tokovinin et al.(2001a)]{Tokovinin01a} Tokivinin, A., Le
%  Louarn, M., Viard, E., et al. 2001, A\&A, 378, 710
\bibitem[Tokovinin et al.(2001)]{Tokovinin01} Tokovinin, A. \&
  Viard, E. 2001, JOSA, 18, 873
\bibitem[Wright et al.(2007)]{Wright07} Wright, S.A., Larkin, J.E.,
  Barczys, M., et al. 2007, ApJ, 658, 78
\bibitem[Yang et al.(2008)]{Yang07} Yang, Y., Flores, H., Hammer, F.,
  et al. 2008, A\&A, 477, 789
\bibitem[Yoshida et al.(2006)]{Yoshida06} Yoshida, M., Shimasaku, K.,
  Kashikawa, N., et al. 2006, ApJ, 653, 988


\end{thebibliography}
\end{document}